\documentclass{aa}  

\usepackage{graphicx}
\usepackage{txfonts}
\usepackage[dvipsnames]{xcolor}
\usepackage[figuresright]{rotating}
\begin{document}

   \title{The Alfvénic nature of chromospheric swirls}

   \author{Andrea~Francesco~Battaglia\inst{1,2,3}
          \and
          José~Roberto~Canivete Cuissa\inst{1,4}
          \and
          Flavio~Calvo\inst{5}
          \and
          Aleksi~Antoine~Bossart\inst{6}
          \and
          Oskar~Steiner\inst{1,7}
          }

   \institute{Istituto Ricerche Solari Locarno (IRSOL),
	         Via Patocchi 57 -- Prato Pernice, 6605 Locarno-Monti, Switzerland\\
              \email{andrea-battaglia@ethz.ch}
         \and
             Institute for Data Science (I4DS), STIX for Solar Orbiter Group, University of Applied Sciences Northwestern Switzerland (FHNW), 5210 Windisch, Switzerland
         \and
             Institute for Particle Physics and Astrophysics (IPA), Solar Astrophysics Group, Swiss Federal Institute of Technology in Zurich (ETHZ), 8039 Zurich, Switzerland
         \and
             Center for Theoretical Astrophysics and Cosmology, Institute for Computational Science (ICS),\\University of Zurich, Winterthurerstrasse 190, 8057 Z{\"u}rich, Switzerland
         \and
             Institute for Solar Physics, Stockholm University, 10691 Stockholm, Sweden
         \and
             Laboratory of Wave Engineering, École Polytechnique Fédérale de Lausanne (EPFL), 1015 Lausanne, Switzerland
         \and
             Leibniz-Institut f\"ur Sonnenphysik (KIS), 
	         Sch\"oneckstrasse 6, 79104 Freiburg i.Br., Germany
             }

   \date{Received 11 December 2020 / Accepted 2 March 2021}

    \abstract
    {Observations show that small-scale vortical plasma motions are ubiquitous in the quiet solar atmosphere. They have received increasing attention in recent years because they are a viable candidate mechanism for the heating of the outer solar atmospheric layers. However, the true nature and the origin of these swirls, and their effective role in the energy transport, are still unclear.}
    {We investigate the evolution and origin of chromospheric swirls by analyzing numerical simulations of the quiet solar atmosphere. In particular, we are interested in finding their relation with magnetic field perturbations and in the processes driving their evolution.}
    { The radiative magnetohydrodynamic code CO$5$BOLD is used to perform realistic numerical simulations of a small portion of the solar atmosphere, ranging from the top layers of the convection zone to the middle chromosphere. For the analysis, the swirling strength criterion and its evolution equation are applied in order to identify vortical motions and to study their dynamics. As a new criterion, we introduce the magnetic swirling strength, which allows us to recognize torsional perturbations in the magnetic field.}
    { We find a strong correlation between swirling strength and magnetic swirling strength, in particular in intense magnetic flux concentrations, which suggests a tight relation between vortical motions and torsional magnetic field perturbations. Furthermore, we find that swirls propagate upward with the local Alfvén speed as unidirectional swirls driven by magnetic tension forces alone. In the photosphere and low chromosphere, the rotation of the plasma co-occurs with a twist in the upwardly directed magnetic field that is in the opposite direction of the plasma flow. All together, these are clear characteristics of torsional Alfvén waves. Yet, the Alfvén wave is not oscillatory but takes the form of a unidirectional pulse. The novelty of the present work is that these Alfvén pulses naturally emerge from realistic numerical simulations of the solar atmosphere. We also find indications of an imbalance between the hydrodynamic and magnetohydrodynamic baroclinic effects being at the origin of the swirls. At the base of the chromosphere, we find a mean net upwardly directed Poynting flux of $12.8\pm6.5\,{\rm kW}\,{\rm m}^{-2}$, which is mainly due to swirling motions. This energy flux is mostly associated with large and complex swirling structures, which we interpret as the superposition of various small-scale vortices. }
    {We conclude that the ubiquitous swirling events observed in numerical simulations are tightly correlated with perturbations of the magnetic field. At photospheric and chromospheric levels, they form Alfvén pulses that propagate upward and may contribute to chromospheric heating.}
    

    \maketitle
   
%
%%%%%%%%%%%%%%%%%%%%%%%%%%%%%%%%%%%%%%%%%%%%%%%%%%%%%%%%%%%%%%%%%%%%%%%%%%%%%%%%%
%%%%%%%%%%%%%%%%%%%%%%%%%%%%%%%%%%%%%%%%%%%%%%%%%%%%%%%%%%%%%%%%%%%%%%%%%%%%%%%%%
%

\section{Introduction}
    Mechanisms for heating the upper solar atmosphere and corona, be they magnetohydrodynamical (MHD) waves or magnetic reconnection, generally rely on the transverse motion of magnetic footpoints in the photosphere. Excitations of kink tube waves or compressive fast modes are often considered to arise from ``granular buffeting,'' by which photospheric magnetic flux concentrations are rapidly and stochastically moved in a sideward direction. Another option is given by turbulent motions in the convection zone, which generate a whole spectrum of transversal excitations. On the other hand, heating by magnetic reconnection is commonly considered to be induced by transverse motions of magnetic footpoints. These can lead to the twisting and braiding of magnetic flux tubes forming dissipative electric current sheets. 
    
    Different from transverse excitations, vortical 
    motions of the magnetic footpoints have been less in the focus of research, mainly 
    for lack of observational support. In the past, vortical motions and torsional Alfv\'en waves were of rather hypothetical nature, often serving theoretical models for the driving of spicules
    \citep[e.g.,][]{1975SoPh...42...79S, % Stenflo (1975)
    1981SoPh...70...25H, % Hollweg (1981)
    2011ApJ...740L..46F}. % Fedun et al. (2011)
    %2015ApJ...808....5M}. % Murawski et al. (2015)
    This state of affairs has drastically changed over the past decade.
    
    In the photosphere, swirls of granular scale were detected by tracking magnetic bright points (BPs) in the intergranular space \citep{2008ApJ...687L.131B, % Bonet et al. (2008)
    2010A&A...513L...6B,  % Balmaceda et al. (2010)
    2011A&A...531L...9M} % Manso Sainz et al. (2011)
    and by the means of local correlation tracking (LCT)  \citep{2010ApJ...723L.139B, % Bonet et al., 2010
    2011MNRAS.416..148V, % Vargaz Dominguez et al., 2011
    2017ApJS..229...14R, % Requerey et al. (2017)
    2018A&A...610A..84R}. % Requerey et al. (2018)
    In the chromosphere, vortical motions with a typical size of about $2\,\mathrm{arcsec}$ were observed by \citet{2009A&A...507L...9W} % Wedemeyer-Bohm & vad der Rouppe Voort, 2009
    and \citet{2012Natur.486..505W} % Wedemeyer et al.,2012
    using \ion{Ca}{ii} 854\,nm narrowband filtergrams obtained with the CRisp Imaging SpectroPolarimeter (CRISP) instrument of the Swedish 1-$\mathrm{m}$ Solar Telescope (SST). Moreover, these chromospheric swirls were colocated with magnetic BPs in the photosphere, suggesting a magnetic origin. 
    Higher up, \citet{2012Natur.486..505W} % Wedemeyer et al.,2012
    detected imprints of chromospheric swirls in the transition region and the corona in the spectral lines of \ion{He}{ii} $30.4\,\mathrm{nm}$ and \ion{Fe}{ix} $17.1\,\mathrm{nm}$, respectively, using recordings of the Atmospheric Imaging Assembly (AIA) instrument of NASA's space-based Solar Dynamics Observatory (SDO). These observations suggested that chromospheric swirls are the observational signatures of rotating coherent magnetic structures, which reach from the convection zone to the outer layers of the atmosphere.
    
    \citet{2013ApJ...768...17M} % Morton et al. (2013)
    found the chromospheric counterpart of photospheric swirls in the form of a quasi-periodic torsional motion, in time series of H$\alpha$ narrowband filtergrams obtained with the ROSA (Rapid Oscillations in the Solar Atmosphere) instrument attached to the Dunn Solar Telescope.
    \cite{2016A&A...586A..25P} % Park et al (2016)
    detected swirls consisting of spiral arms in H$\alpha$ filtergrams taken with CRISP at SST, which coexisted with strong upflows in the upper chromosphere as seen from \ion{Mg}{ii} k line Dopplergrams obtained with the space-based Interface Region Imaging Spectrograph (IRIS).
    More recently, a $1.7\,\mathrm{h}$ persistent vortex flow of $6\,\mathrm{arcsec}$ diameter was observed by \citet{2018A&A...618A..51T, 2019A&A...623A.160T} % Tziotziou et al., 2018 & 2019
    with CRISP at SST in the cores of the H$\alpha$ and \ion{Ca}{ii} $854.2\,\mathrm{nm}$ lines, while no BPs were observed in the line wings of H$\alpha$ and \ion{Ca}{ii}.
    Finally, \cite{2019ApJ...881...83S} % Shetye et al. (2019)
    found, from spectral imaging in the lines of H$\alpha$ and \ion{Ca}{ii} $854\,{\rm nm}$ along with polarimetry in \ion{Fe}{i} $630.2\,{\rm nm}$, that the rotation of magnetic flux concentrations in the photosphere matches the chromospheric swirl. However, they reported that they could not determine whether a swirl is a gradual response to the photospheric motion or an actual propagating Alfvénic wave.
    
    Alfv\'en waves were theoretically predicted by \citet{1942Natur.150..405A}, but their detection in the Sun, which provides preferential conditions for their existence, proved to be a difficult task. Nevertheless, over the past two decades, multiple observations have revealed their presence and propagation in the solar atmosphere and corona 
    \citep[e.g.,][]{2007Sci...317.1192T, % Tomczyk & McIntosh (2007)
    2011ApJ...736L..24O, % Okamoto & De Pontieu (2011)
    2007Sci...318.1577O}. % Okamoto et al. (2007)
    \citet{2009Sci...323.1582J} % Jess et al., 2009
    found signatures of propagating torsional Alfv\'en waves in the photosphere in oscillatory phenomena associated with a conglomeration of magnetic BPs.
    \citet{2012ApJ...752L..12D} % De Pontieu et al. (2012)
    reported rotational motion in spicules, which they interpreted to be torsional Alfv\'en waves, and \citet{2017NatSR...743147S} %Srivastava et al., 2017
    detected ubiquitous high frequency (${\sim}12\,\text{-}\,42\,\mathrm{mHz}$) torsional motions in thin spicular-type structures in the chromosphere that resemble torsional Alfv\'en waves.
    \citet{2019NatCo..10.3504L} % Liu et al., 2019
    called attention to ubiquitous torsional Alfv\'en pulses by correlating photospheric and chromospheric swirls.
    Moreover, \citet{2018NatPh..14..480G} provided observational evidence of Alfv\'en waves heating  chromospheric plasma in a sunspot umbra through the formation of shocks. 

    Vortex flows and their connection with Alfv\'en waves have also been extensively studied with MHD numerical simulations. Small-scale swirls appear regularly in simulations of the solar surface convection and atmosphere \citep[see, e.g.,][]{
    1998ApJ...499..914S,  % Stein & Nordlund (1998)
    2004RvMA...17...69V,  % Vögler (2004)
    2011A&A...526A...5S,  % Shelyag et al. (2011)
    2011A&A...533A.126M,  % Moll et. al. (2011)
    2012A&A...541A..68M,  % Moll et. al. (2012)
    2012ASPC..456....3S,  % Steiner & Rezaei (2012)
    2013ApJ...770...37K,  % Kitiashvili et al. (2013)
    %2014PASJ...66S..10W,  % Wedemeyer & Steiner (2014)
    2016A&A...596A..43C,  % Calco et al. (2016)
    2019A&A...632A..97L,  % Liu et al. (2019)
    2020ApJ...894L..17Y}. %Yadav et al. (2020)
    \citet{2013ApJ...776L...4S} % Shelyag et al. 2013
    identified apparent vortex-like motions in magnetic flux tubes of the photosphere as torsional Alfv\'en waves by analyzing time-distance diagrams and plotting particle tracks.
    \citet{2014PASJ...66S..10W} % Wedemeyer et al., 2014
    put forward a model consisting of two vortex systems, where the upper chromospheric and photospheric swirling plasma is tightly coupled to the ``frozen-in'' magnetic field lines, which have their footpoints within the lower  intergranular vortex flow. 
    This system of vortical structures was dubbed a ``magnetic tornado'' \citep{2012Natur.486..505W}. % Wedemeyer et al., 2012
    \citet{2011AnGeo..29.1029F}, % Fedun et al. (2011)
    \citet{2012ApJ...755...18V}, % Vigeesh et al. (2012)
    and \citet{2015A&A...577A.126M} % Murawski et al., 2015
    studied the generation, propagation, and energy transfer of torsional Alfv\'en waves in modeled solar magnetic flux tubes, and \citet{2019NatCo..10.3504L} showed that azimuthal perturbations of the magnetic field in the upper photosphere can generate Alfv\'en pulses, which carry the information of the rotational motion into the chromosphere. However, the exact mechanism at the origin of the swirls in realistic numerical simulations is still unclear.
    
    A particularly interesting aspect of these swirls is that they possibly generate torsional Alfv\'en waves that propagate upward and provide an efficient mechanism for chromospheric and coronal heating. From numerical simulations, \citet{2012Natur.486..505W} % Wedemeyer et al.,2012
    estimated a net vertical Poynting flux associated with swirling motions of $440\,\mathrm{W\, m^{-2}}$ at the interface between the chromosphere and the corona, while \citet{2019NatCo..10.3504L} % Liu et al., 2019
    estimate a minimal nonthermal energy flux of $33\,\text{-}\,131\,\mathrm{W\,m^{-2}}$ in the middle chromosphere.

    In this paper, we study in detail ``magnetic swirls,'' that is, the relation between swirls, magnetic field perturbations, and Alfv\'en waves.  
    In particular, we investigate the MHD processes that lead to the propagation of vortical motions from the photosphere to the chromosphere, where they appear as chromospheric swirls. For that purpose, we analyze realistic radiative MHD numerical simulations carried out with the CO$5$BOLD code 
    \citep{2012JCoPh.231..919F} % Freytag et al., 2012
    by using the swirling strength criterion and its evolution equation 
    \citep{1999JFM...387..353Z, % Zhou et al., (1999)
    2020A&A...639A.118C}. % Canivete & Steiner (2020)

    The paper is organized as follows: Section \,\ref{sec:theory} reviews a number of theoretical concepts that are used for the analysis.  Section\,\ref{sec:numerics} gives details on the numerical simulations and Sect.\,\ref{sec:results}  presents the analysis and discusses the results. 
    A summary and conclusions are given in Sect.\,\ref{sec:conclusions}.
    
%
%%%%%%%%%%%%%%%%%%%%%%%%%%%%%%%%%%%%%%%%%%%%%%%%%%%%%%%%%%%%%%%%%%%%%%%%%%%%%%%%%
%%%%%%%%%%%%%%%%%%%%%%%%%%%%%%%%%%%%%%%%%%%%%%%%%%%%%%%%%%%%%%%%%%%%%%%%%%%%%%%%%
%

\section{Theoretical background}\label{sec:theory}
    
    This section reviews a few theoretical concepts that are used for the analysis of the simulations. These concern properties of Alfv\'en waves, the computation of energy fluxes and the concept of swirling strength. Moreover, we introduce a new quantity, the ``magnetic swirling strength,'' which is used for detecting twists in magnetic fields.
    
    %
    %%%%%%%%%%%%%%%%%%%%%%%%%%%%%%%%%%%%%%%%%%%%%%%%%%%%%%%%%%%%%%%%%%%%%%%%%%%%%
    %
    
    \subsection{Alfvén waves \label{subsec:AlfvénWaves}}
    
    Alfvén waves are perturbations in the plasma with the magnetic tension acting as a restoring force. The perturbations are transverse to the propagation direction and magnetic field lines; therefore, one speaks of shear or torsional Alfv\'en waves. 
    When the perturbation propagates across the magnetic field, causing compression and rarefaction of the magnetic field and plasma, one speaks of compressional Alfv\'en waves or fast mode waves \citep[see, e.g.,][]{2014masu.book.....P}. % Priest, 2014.
    For the purposes of this paper, we focus solely on the former. 

\begin{figure}
        \centering
        \includegraphics[width=0.6\hsize]{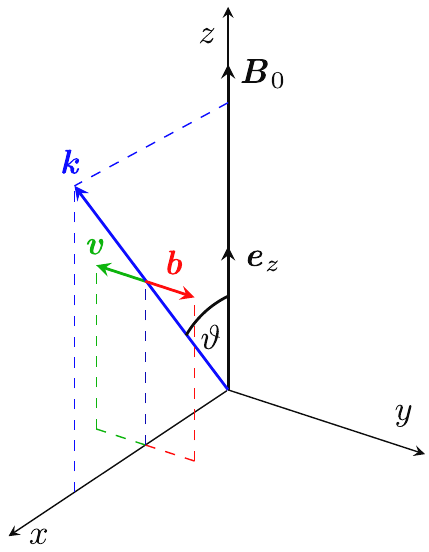}
        \caption{Torsional Alfv\'en wave propagating in the direction $\boldsymbol{k}$. 
        Perturbations in the magnetic field, $\boldsymbol{b}$, and in the plasma velocity field, $\boldsymbol{v}$, are normal to the plane spanned by $\boldsymbol{k}$ and $\boldsymbol{B}_0$.
        }
        \label{fig:AlfvenWavesScheme}
    \end{figure}

    We consider with Fig.\,\ref{fig:AlfvenWavesScheme} a static, incompressible plasma in the ideal MHD approximation, with constant density $\rho$ and uniform and vertical magnetic field $\boldsymbol{B}_0 = B_0 \boldsymbol{e}_z$. Moreover, we suppose that the magnetic field dominates the equilibrium so that the hydrodynamical pressure and gravity can be neglected. Although these conditions are not satisfied in the solar atmosphere, this abstraction still serves to describe some basic behavior of the Alfv\'en wave propagation in the solar photosphere and chromosphere.
    Introducing the magnetic field perturbation with $\boldsymbol{B} = \boldsymbol{B}_0 + \boldsymbol{b}$, where $|\boldsymbol{b}| \ll |\boldsymbol{B}_0|$, the velocity perturbation $\boldsymbol{v}$, and linearizing the system of MHD equations using incompressibility
    \citep[see, e.g.,][Chap. 4]{2014masu.book.....P}, % Priest, 2014
    yields two important characteristics of Alfv\'en wave propagation.

    The first one relates the velocity perturbation to the magnetic field perturbation 
    \begin{equation} \label{eqn:PerturbationRelation}
        \boldsymbol{v} = -\dfrac{\omega\boldsymbol{b}}{\boldsymbol{k} \cdot \boldsymbol{B}_0} =  \left\{\begin{array}{r@{\;\mathrm{\quad if}\;}l}
        -(v_A/B_0)\,\boldsymbol{b} & 0\le \vartheta < \pi/2\,, \\[1.0ex]
        (v_A/B_0)\,\boldsymbol{b} & \pi/2 < \vartheta \le \pi\,,
        \end{array}\right.
    \end{equation}
    where  $\omega$ is the angular frequency of the plane wave, $\boldsymbol{k}$ the wave-vector indicating the propagation direction,
    $v_{\rm A} = B_0 / \sqrt{4 \pi \rho}$ is the Alfvén speed, and $\vartheta$ is the angle between $\boldsymbol{e}_z$ and $\boldsymbol{k}$. We note that, if $\vartheta < \pi/2$, the perturbations $\boldsymbol{v}$ and $\boldsymbol{b}$ are anti-parallel, while they are parallel if $\vartheta > \pi/2$.
    In the following, for simplicity, we only consider the case in which the wave travels in the direction of the magnetic field, that is, $0 \le\vartheta < \pi/2$. Furthermore, one can prove that both the magnetic field and the velocity perturbation are azimuthal with respect to the static magnetic field, $\boldsymbol{B}_0$, and the wave-vector $\boldsymbol{k}$ can point in any direction as depicted in Fig.\,\ref{fig:AlfvenWavesScheme}. This implies that the linearized version of the magnetic pressure is null because $p_{\rm m} = \boldsymbol{b}\cdot\boldsymbol{B}_0/8\pi = 0$. Since plasma pressure and gravity are not considered, the only driving force of torsional Alfv\'en waves is the magnetic tension. 

    The second result is the dispersion relation
    \begin{equation}
        \omega^2 = v_A^2 (\boldsymbol{k}\cdot\boldsymbol{e}_z)^2\,. \nonumber
    \end{equation}
    From this equation, it is possible to obtain the group velocity of an Alfvén wave packet, which corresponds to the velocity at which the energy is transmitted, 
    \begin{equation}
        \boldsymbol{v}_{\text{group}} = \dfrac{\partial \omega}{\partial \boldsymbol{k}} = v_A \boldsymbol{e}_z 
        \quad\mathrm{if}\; 0\le \vartheta < \pi/2\,.
    \end{equation}
    It is equivalent to the Alfvén speed and energy is carried along the equilibrium magnetic field. 
       
    %
    %%%%%%%%%%%%%%%%%%%%%%%%%%%%%%%%%%%%%%%%%%%%%%%%%%%%%%%%%%%%%%%%%%%%%%%%%%%%%
    %
    
    \subsection{Energy fluxes \label{subsec:energyflux}}
        
    Alfvén waves can contribute to the energy transport in the solar atmosphere. To study this aspect, the MHD Poynting flux vector is employed,
    \begin{equation}
        \boldsymbol{S} = \dfrac{1}{4\pi} \boldsymbol{B} \times (\boldsymbol{v} \times \boldsymbol{B}) = \dfrac{1}{4\pi} (\boldsymbol{B} \cdot \boldsymbol{B})\,\boldsymbol{v} - \dfrac{1}{4\pi} (\boldsymbol{B} \cdot \boldsymbol{v})\, \boldsymbol{B}\,. \label{eq:poytingvector}
    \end{equation}
    Since we are interested in the vertical energy flux, the $z$-component of Eq.\,(\ref{eq:poytingvector}) is expanded following 
    \citet{2012ApJ...753L..22S}, % Shelyag et al., 2012, 
    \begin{equation}
        S_z = \underbrace{\dfrac{1}{4\pi} v_z (B_x^2 + B_y^2)}_{S_z^{\rm v}}\, - \underbrace{\dfrac{1}{4\pi} B_z (v_x B_x + v_y B_y)}_{S_z^{\rm h}}, \label{eqn:z-Poynting}
    \end{equation}
    where the first term, $S_z^{\rm v}$, is related to vertical motions of horizontal magnetic field and the second term, $S_z^{\rm h}$, corresponds to the vertical flux generated by horizontal motions of magnetized plasma. 

    The contribution of the swirls to the mean Poynting flux over a given field of view of area $A_{\rm FOV}$ is then given by
    \begin{equation}
        \bar{S}_z = S_z \bar{N}_{\rm s} \frac{\pi\bar{r}_{\rm s}^2}{A_{\rm FOV}}\,, \label{eq:averagesflux}     
    \end{equation}
    where $S_z$ is the mean contribution from a single swirl, $\bar{N}_{\rm s}$ is the average number of swirls that can be observed at any time in the given field of view, be it of a simulation or an observation, and $\bar{r}_{\rm s}$ is the average swirl radius. 

    A swirl can also produce a mechanical energy flux, which is given by
    \begin{equation}
        F_{z}^{\rm r} = \frac{1}{A_{\rm s}}\iint_{A_{\rm s}}\mathrm{d}x\mathrm{d}y\, \frac{1}{2} \rho v_{\rm r}^2 v_{\rm p} \,, \label{eq:mechflux_vh} 
    \end{equation}
    where the integration is taken over the assumed circular area $A_{\rm s}$ associated with the swirl, $\rho = \rho(x,y)$ is the density of the plasma, and $v_{\rm r} = v_{\rm r}(x,y)$ is the rotational velocity. Furthermore, $v_{\rm p} = v_{\rm p}(x,y)$ is the propagation speed of the swirl in the vertical direction. For simplicity, we assume $A_{\rm s}$ to lie in a horizontal plane and $v_{\rm r}$ to be the velocity projected into this plane. Assuming that the flow rotates around a common axis, we can express the rotational velocity in terms of the vertical component of the swirling strength vector as $v_{\rm r} = \frac{1}{2}r \lambda_z$ (see Sect.\,\ref{subsec:vortex_detection}), hence
    \begin{equation}
        F_{z}^{\lambda} = \frac{1}{8 A_{\rm s}}\iint_{A_{\rm s}} \mathrm{d}x\mathrm{d}y\, \rho r^2 \lambda_z^2 v_{\rm p} \,, \label{eq:mechflux_vlambda}
    \end{equation}
    where $r = \sqrt{x^2 + y^2}$ is the distance from the center of the swirl and $\lambda_z = \lambda_z(x,y)$ is the associated vertical component of the swirling strength vector.
    
    The contribution of swirling motions to the mean mechanical energy flux $\bar{F}_z$ is then
    \begin{equation}
        \bar{F}_z = F_z^{{\rm r},\lambda} \bar{N}_{\rm s} \frac{\pi\bar{r}_{\rm s}^2}{A_{\rm FOV}}\,,
        \label{eq:averagemflux}     
    \end{equation}
    where $F^{{\rm r},\lambda}_z$
    is the mean mechanical flux density of a single swirl computed with either Eq.\,(\ref{eq:mechflux_vh}) or Eq.\,(\ref{eq:mechflux_vlambda}).
    We notice that $S_z$, $F_z$, $\bar{F}_z$ and $\bar{S}_z$ are energy flux densities.
        
    %
    %%%%%%%%%%%%%%%%%%%%%%%%%%%%%%%%%%%%%%%%%%%%%%%%%%%%%%%%%%%%%%%%%%%%%%%%%%%%%
    %

    \subsection{Swirling strength} \label{subsec:vortex_detection}
    
    A vortex or swirl can be intuitively described as the rotation of fluid parcels around a common axis. Despite this simple concept, a rigorous mathematical definition is still an open issue in fluid mechanics \citep[see, e.g.,][]{Kolar2007,2019JHyDy..31..205L}. %Kolar (2007), Liu et al (2019)
    The vorticity is the classical quantity to describe rotational flows and it has been adopted to investigate vortex flows in numerical simulations of the solar convection zone \citep[see, e.g.,][]{1997A&A...328..229N} %Nordlund et al. (1997) 
    and atmosphere \citep[see, e.g.,][]{2011A&A...526A...5S, 2013ApJ...776L...4S}. %Shelyag et al. (2011), %Shelyag et al. (2013)
    However, \citet{1995JFM...285...69J} %Jeong et Hussein, 1995
    showed that the vorticity is not a suitable tool for vortex identification since it cannot distinguish between non-rotational shear flows and actual vortices. This can lead to misidentifications in a dynamical and turbulent system such as the solar atmosphere. Therefore, we decided to employ the swirling strength criterion, which is not affected by shears  \citep{1999JFM...387..353Z}. %Zhou et al, 1999

    The swirling strength is based on the eigenanalysis of the velocity gradient tensor $\mathcal{U}_{ij} \coloneqq \partial_j v_i$, that is, the Jacobian matrix of the local velocity field. If a vortex is present in the flow, the velocity gradient tensor can be locally diagonalized and it will exhibit one real and a pair of complex conjugated eigenvalues
    \begin{eqnarray}
      \mathcal{U} =  
        \underbrace{\vphantom{\begin{bmatrix}
    		                  \lambda_{\rm r} & 0 &  0\\
    		                  0 & \lambda_{\rm +}& 0 \\
    		                  0 & 0 & \lambda_{\rm -}  
    		                  \end{bmatrix}} 
        \left[ \boldsymbol{u}_{\rm r}, \boldsymbol{u}_{\rm +}, \boldsymbol{u}_{\rm -}\right]
        }_{\textstyle\mathcal{P}}
    	\underbrace{%
    	\begin{bmatrix}
          \lambda_{\rm r} & 0 &  0\\
          0 & \lambda_{\rm +}& 0 \\
          0 & 0 & \lambda_{\rm -}  
        \end{bmatrix}}_{\textstyle\Lambda}
        \underbrace{\vphantom{\begin{bmatrix}
    		                  \lambda_{\rm r} & 0 &  0\\
    		                  0 & \lambda_{\rm +}& 0 \\
    		                  0 & 0 & \lambda_{\rm -}  
    		                  \end{bmatrix}} 
    	\left[ \boldsymbol{u}_{\rm r}, \boldsymbol{u}_{\rm +}, \boldsymbol{u}_{\rm -}\right]^{-1}
    	}_{\textstyle\mathcal{P}^{-1}}\,,  \label{eq:U_decomposition}
    \end{eqnarray}
    where $\lambda_{\pm} = \lambda_{\rm cr} \pm  \mathrm{i}\lambda_{\rm ci}$ and $\lambda_{\rm r} \in \mathbb{R}$. 
    The swirling strength $\lambda$ is then defined through the imaginary part of the complex eigenvalues, $\lambda_{\rm ci}$. In this paper, we adopt the same convention as in \citet{2020A&A...639A.118C} % Canivete & Steiner, 2020
    and define the swirling strength as $\lambda \coloneqq 2\lambda_{\rm ci}$. For a rigid body rotation, the period of revolution  $T$ of the flow is then computed as $T = 4\pi/\lambda$ and the vorticity $\omega \coloneqq |\boldsymbol{\nabla}\times\boldsymbol{v}|$ equals the swirling strength $\lambda$. 
    Furthermore, the rotation axis and the orientation of the vortex can be inferred from the normalized eigenvector associated with the real eigenvalue, $\boldsymbol{u}_{\rm r}$. Hence, it is possible to define a swirling vector, $\boldsymbol{\lambda} \coloneqq \lambda \boldsymbol{u}_{\rm r}$, which carries the necessary information to characterize the vortex in three dimensions. For a detailed review of the swirling strength, the reader can refer to \citet{2020A&A...639A.118C}. % Canivete & Steiner, 2020

    The swirling strength is invariant under Galilean transformations and can be used in the context of compressible \mbox{(magneto-)hydrodynamics} \citep{2009AIAAJ..47..473K}. % Kolar, 2009
    This criterion has already been successfully applied in studies regarding vortex flows in the solar atmosphere \citep[see][]{2011A&A...533A.126M, 2017A&A...601A.135K, 2020ApJ...894L..17Y}. %Moll et al.,2011; Kato & Wedemeyer, 2017; Yadav et al, 2020
    Furthermore, \citet{2020A&A...639A.118C} % Canivete & Steiner, 2020
    derived the evolution equation for the swirling strength. A similar equation was already known for the vorticity \citep[see, e.g.,][]{2011A&A...526A...5S} % Shelyag et al, 2011
    but, in this case, it can also be biased by shear flows. Therefore, the swirling equation is a valuable tool for the investigation of the physical processes responsible for the dynamics of vortices. It can be derived from a general momentum equation of \mbox{(magneto-)hydrodynamics}: For the specific case of ideal MHD, it can be formulated as
    \begin{align}
        \frac{{\rm D} \lambda }{{\rm D}t} =&  
        \vphantom{\left\{ \mathcal{P}^{-1} \left[ \nabla \bigg(  \frac{1}{\rho}\nabla p_{\rm g} \bigg) \right] \mathcal{P} \right\}}
        - 2\lambda\lambda_{\rm cr} & T^1 & \nonumber \\
        & - 2\rm{Im}\left\{ \mathcal{P}^{-1} \left[ \nabla \otimes \bigg(  \frac{1}{\rho}\nabla \textit{p}_{\rm g} \bigg) \right] \mathcal{P} \right\}_{22} & T^2 & \nonumber \\
        & - 2\rm{Im}\left\{ \mathcal{P}^{-1} \left[ \nabla \otimes \bigg(  \frac{1}{\rho} \nabla  \textit{p}_{\rm m} \bigg) -  \bigg(\nabla\frac{1}{\rho} \bigg)\otimes \frac{ (\boldsymbol{B}\cdot\nabla) \boldsymbol{B}}{4\pi}  \right] \mathcal{P} \right\}_{22} & T^3 & \nonumber \\
        & + 2\rm{Im}\left\{ \mathcal{P}^{-1} \left[ \frac{1}{\rho} \nabla \otimes  \frac{( \boldsymbol{B} \cdot \nabla ) \boldsymbol{B}}{4\pi}  \right] \mathcal{P} \right\}_{22} & T^4 & \nonumber \\
        & \vphantom{\left\{ \mathcal{P}^{-1} \left[ \nabla \bigg(  \frac{1}{\rho}\nabla p_{\rm g} \bigg) \right] \mathcal{P} \right\}} 
        - 2\rm{Im}\left\{ \mathcal{P}^{-1} \bigg[ \nabla \otimes \Big( \nabla \Phi \Big) \bigg] \mathcal{P} \right\}_{22}\,, & T^5 
        &    \label{eq:swirling_eq}
    \end{align} 
    where $\lambda_{\rm cr}$ is the real component of the complex eigenvalues of $\mathcal{U}$, $\mathcal{P}$ and $\mathcal{P}^{-1}$ are, respectively, the matrix composed of the eigenvectors of $\mathcal{U}$ and its inverse as shown in Eq.\,(\ref{eq:U_decomposition}), $p_{\rm g}$ is the atmospheric gas pressure, $p_{\rm m} = B^2/8\pi$ represents the pressure owing to the magnetic field, and $\Phi$ corresponds to a potential of conservative forces. We notice that the terms between curly brackets are matrices since the symbol $\otimes$ denotes the tensor product between two vectors. We are only interested in the $(2,2)$ component of the resulting matricial multiplication. 
    Finally, the material derivative on the left-hand side of Eq.\,(\ref{eq:swirling_eq}) is defined as
    \begin{equation}
        \frac{{\rm D} \lambda }{{\rm D}t} = \frac{\partial \lambda}{\partial t} + \left(\boldsymbol{v}\cdot\nabla\right) \lambda\,, \nonumber
    \end{equation}
    where $\partial \lambda/\partial t$ defines the local production of swirling strength, while $\left(\boldsymbol{v}\cdot\nabla\right) \lambda$ accounts for the swirling strength advected with the plasma.
    
    Following \citet{2020A&A...639A.118C}, % Canivete & Steiner, 2020 
    we give a physical interpretation for each one of the terms appearing in Eq.\,(\ref{eq:swirling_eq}). The first term, $T^1$, is a stretching term, while $T^2$ and $T^3$ are related to the baroclinic effects: The former is a pure hydrodynamical process and the latter is due to magnetic fields. Then, $T^4$ is associated with magnetic tension effects and $T^5$ describes the generation of swirling strength by conservative forces. 
    
    For the statistical analyses, plots in Sect.~\ref{sec:results}, and the appendices, a swirling strength threshold is applied to ease the interpretation and reduce noise \citep{1999JFM...387..353Z}. % Zhou et al. 1999
    It is also motivated by the request that swirls perform a significant fraction of a full rotation over the time span of the simulation. 
    Therefore, swirls having a period longer than $10\,\mathrm{min}$, that is, $\lambda < 2.09\,\times\,10^{-2}\,\mathrm{Hz}$, are discarded.
    %
    %%%%%%%%%%%%%%%%%%%%%%%%%%%%%%%%%%%%%%%%%%%%%%%%%%%%%%%%%%%%%%%%%%%%%%%%%%%%%
    %
    
    %
    \begin{figure*}
        \resizebox{\hsize}{!}{\includegraphics{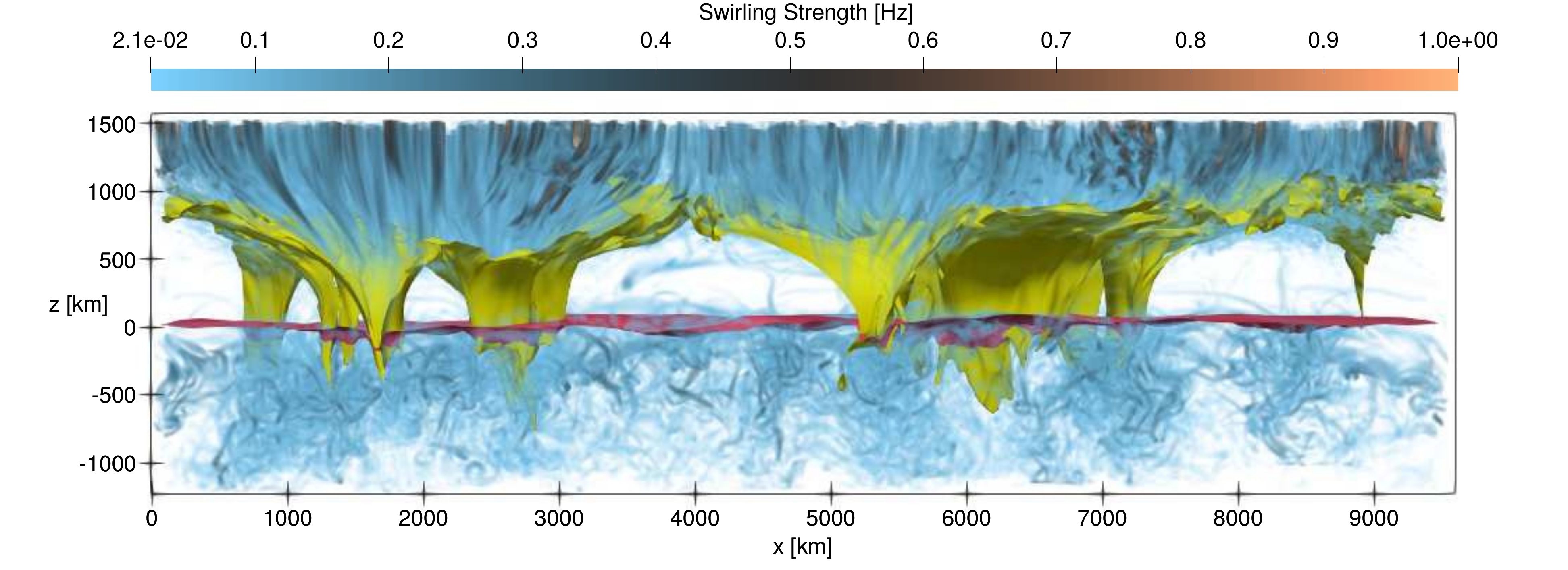}}
        \caption{
        Time instant of the swirling strength $\lambda$ in a slice of dimension $9.6\times2.6\times2.8\,\mathrm{Mm}^3$ of the whole physical domain of the numerical simulation. The red contour indicates the surface of optical depth $\tau_{500} = 1$. The yellow contour corresponds to the isosurface of plasma-$\beta = 1$. Swirls with a period larger than $10\,\mathrm{min}$ ($\lambda < 2.09\,\times\,10^{-2}\,\mathrm{Hz}$) are not shown and the values $\lambda > 1\,\mathrm{Hz}$ are saturated. This rendering has been produced with ParaView \citep{Ahrens+al2005}. %Ahrens et al (2005)
        }
        \label{fig:SwirlsDistribution_TauBeta}
    \end{figure*}

    \subsection{Magnetic swirling strength}
    \label{subsec:magneticswirlingstrength}
    
    Torsional Alfvén waves are characterized by azimuthal perturbations in the vertically directed magnetic field lines. In order to identify this torsion in numerical simulations, we define the criterion of magnetic swirling strength, $\lambda^{B}$. This quantity is derived in the same way as the swirling strength, but the matrix to be diagonalized is the magnetic gradient tensor, $\mathcal{B}_{ij} \coloneqq \partial_j B_i$. Hence, the magnetic swirling strength is defined through the imaginary component of the complex eigenvalues of $\mathcal{B}$, that is, $\lambda^{B} \coloneqq 2\lambda^{B}_{\rm ci}$. Following the same reasoning as for the swirling strength, the magnetic swirling vector $\boldsymbol{\lambda}^{B}=\lambda^{B} \boldsymbol{u}^{B}_{\rm r}$ can be defined, which describes the three-dimensional twisting of the magnetic field. 
    It is worth noticing that the units of the magnetic swirling strength are $[\mathrm{G\,cm^{-1}}]$ and that therefore this quantity does not describe a rotational flow of the magnetic field lines, but rather their twisting around the axis represented by the real eigenvector, $\boldsymbol{u}_{\rm r}^{B}$. Thus, a magnetic field line rotating together with the plasma flow in a rigid body fashion will produce no magnetic swirling strength, while a magnetic field line with an helical structure will present a finite value of magnetic swirling strength. Consequently, this criterion is well suited to detect azimuthal torsions in the magnetic field and, therefore, imprints of torsional Alfvén waves. 

    For torsional Alfvén waves in a static, incompressible plasma dominated by the magnetic field, we can find a simple relation between the swirling strength $\lambda$ and the magnetic swirling strength $\lambda^{B}$. By taking the tensor product between $\nabla$ and Eq.\,(\ref{eqn:PerturbationRelation}) it is possible to obtain
    \begin{equation}
        \mathcal{U} = -\dfrac{1}{\sqrt{4\pi\rho}}\,\mathcal{B} 
        \quad\mathrm{if}\;0\le \vartheta < \pi/2\,,
    \end{equation}
    which can be diagonalized on both sides since $1/\sqrt{4\pi\rho}$ is a real scalar. 
    Given the definitions of $\lambda$ and $\lambda^{B}$, we get
    \begin{equation}
        \lambda = -\dfrac{1}{\sqrt{4\pi\rho}}\,\lambda^{B} \quad\mathrm{if}\; 0\le \vartheta < \pi/2\,,
        \label{eq:relationlambdas}
    \end{equation}
    which simply states that, for a torsional Alfv\'en wave propagating in direction of the magnetic field ($\vartheta < \pi/2$), the swirling strength and the magnetic swirling strength are proportional to each other and have opposite sign (see also Fig.\,\ref{fig:AlfvenWavesScheme}). 

%
%%%%%%%%%%%%%%%%%%%%%%%%%%%%%%%%%%%%%%%%%%%%%%%%%%%%%%%%%%%%%%%%%%%%%%%%%%%%%%%%%
%%%%%%%%%%%%%%%%%%%%%%%%%%%%%%%%%%%%%%%%%%%%%%%%%%%%%%%%%%%%%%%%%%%%%%%%%%%%%%%%%
%
    
\section{Numerical simulation}\label{sec:numerics}
    
    The high-resolution and high-cadence simulations analyzed in this study have been carried out with the radiative MHD code CO$5$BOLD \citep{2012JCoPh.231..919F}, % Freytag et al., 2012
    which solves the coupled system of compressible MHD equations in an external gravitational field and includes nonlocal, frequency-dependent radiative transfer. 

    The simulations are performed on a three-dimensional Cartesian grid of size $9.6\times9.6\times2.8\,\mathrm{Mm^3}$, with a grid cell size of $10\,\mathrm{km}$ in each spatial dimension. The vertical component ranges from about $1300\,\mathrm{km}$ below the optical surface $\tau_{500} = 1$ to $1500\,\mathrm{km}$ above it. Hence, the simulation domain encompasses a small volume near the solar surface, ranging from the surface layers of the convection zone to the middle chromosphere. The gravitational field is uniform and vertical with a value of $\log{(g)} = 4.44$.

    The initial condition of the simulation is given by a previously relaxed hydrodynamical model, to which a uniform and vertical magnetic field of $50\,\mathrm{G}$ was superimposed. The system is then advanced with the MHD module of CO$5$BOLD \citep{2005ESASP.596E..65S} % Schaffenberger et al, 2005
    until relaxation. The lateral boundary conditions are periodic for both plasma and magnetic fields, while the top and bottom ones are open under the condition that the net mass flux at the bottom is zero. An entropy inflow is supplied to maintain an average surface effective temperature of $T_{\rm eff} = 5770\,\mathrm{K}$. The magnetic field is constrained to be vertical at the top and bottom boundaries. More details on the simulation setup are given in 
    \citet[][Sect.\,2]{doi:10.13097/archive-ouverte/unige:115257}, % Calvo, 2018
    in particular run \texttt{d3gt57g44v50fc}. 
    
    For the present study, a time series of $441$ three-dimensional data cubes with a cadence of $2\,\mathrm{s}$, amounting to about $5\,{\rm TB}$ of data, is analyzed. It starts from $t\!=\!5520\,{\rm s}$ of
    \texttt{d3gt57g44v50fc}.
    This relatively high-cadence has proven to be necessary to be able to capture the evolution and the dynamics of the detected vortices. The series spans a period of approximately $15\,{\rm min}$.
    
%
%%%%%%%%%%%%%%%%%%%%%%%%%%%%%%%%%%%%%%%%%%%%%%%%%%%%%%%%%%%%%%%%%%%%%%%%%%%%%%%%%%
%%%%%%%%%%%%%%%%%%%%%%%%%%%%%%%%%%%%%%%%%%%%%%%%%%%%%%%%%%%%%%%%%%%%%%%%%%%%%%%%%%
%

    \begin{figure*}
        \centering
        %{\includegraphics{Figures/Correlation_Spearman_with200.pdf}}
        \resizebox{\hsize}{!}{
        \begin{minipage}[b]{.487\linewidth}
            \includegraphics[width=\hsize]{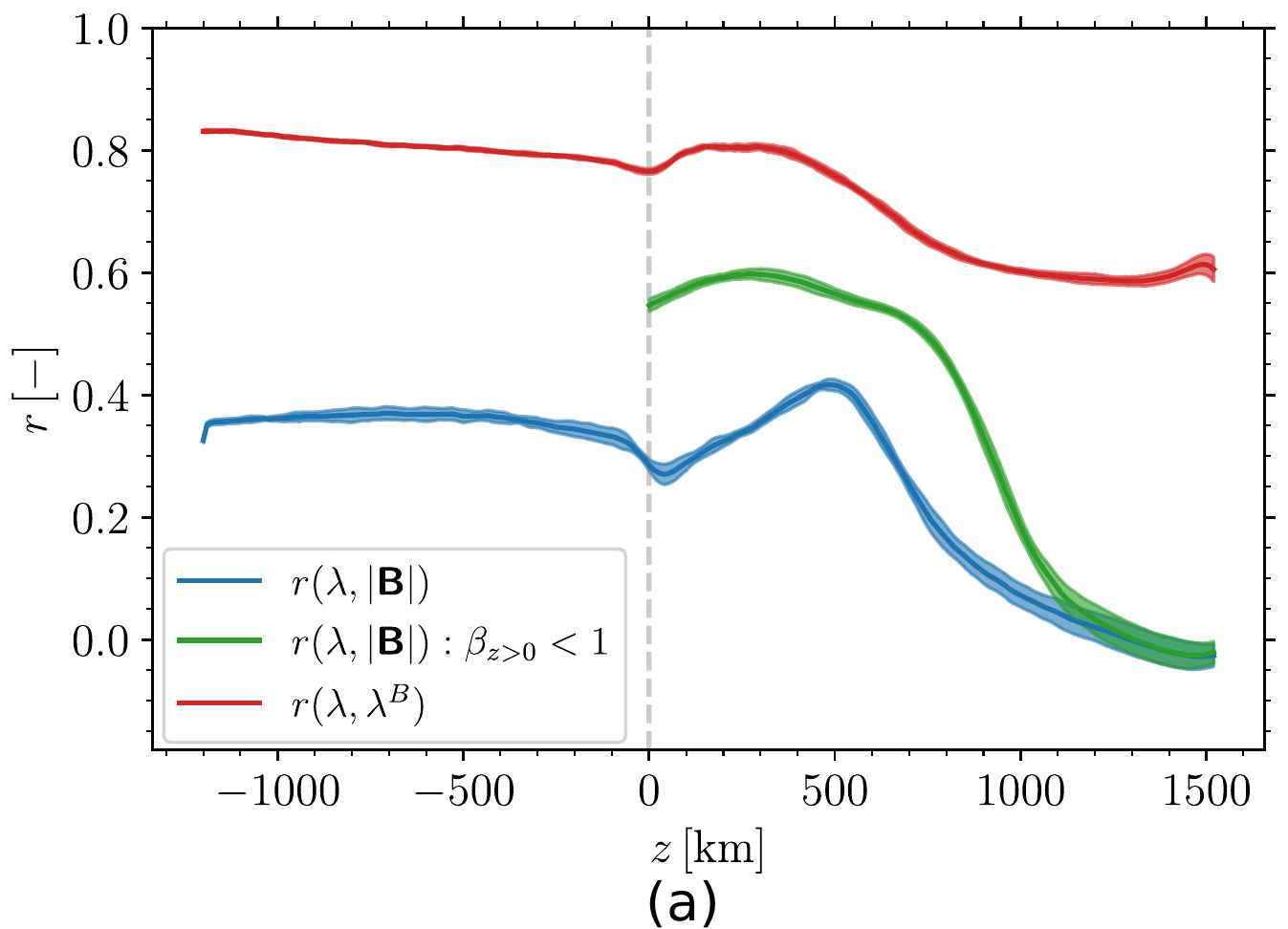}
            %\subcaption{}
            \label{fig:SpearmanR}
        \end{minipage}
        \hspace{0.3cm}
        \begin{minipage}[b]{.518\linewidth}
            \includegraphics[width=\hsize]{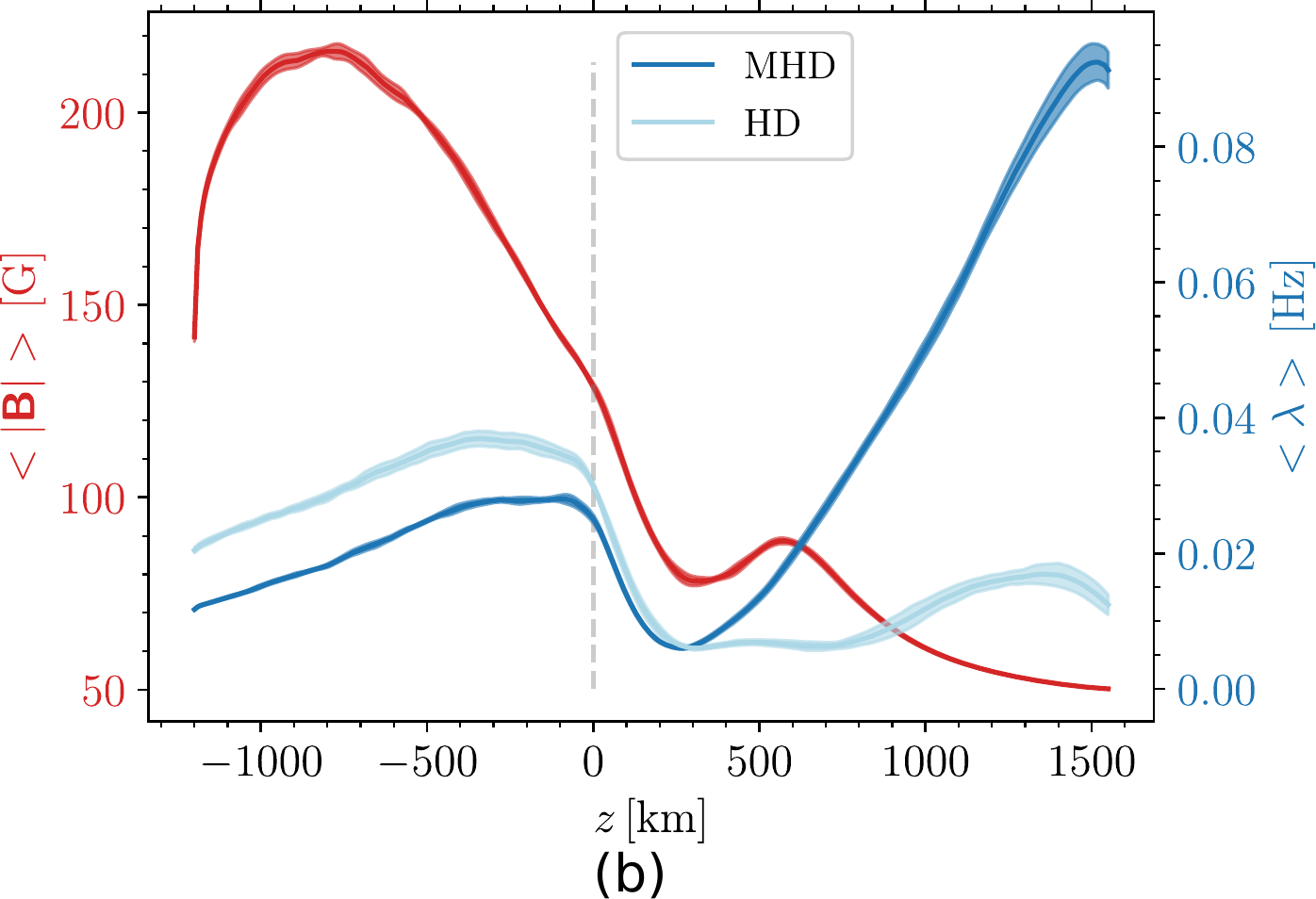}
            %\subcaption{}
            \label{fig:Mean_BandSS}
        \end{minipage}
        }
        \caption{Statistical properties of swirling motions and mean physical quantities as a function of height $z$. (\emph{a}) Spearman's correlation coefficient $r$ between the swirling strength $\lambda$ and the magnetic field strength $|\boldsymbol{B}|$, $r(\lambda, |\boldsymbol{B}|)$ in blue, and between the swirling strength $\lambda$ and the magnetic swirling strength $\lambda^{B}$, $r(\lambda, \lambda^{B})$ in red. The green curve represents $r(\lambda, |\boldsymbol{B}|)$ calculated in regions in which $\beta < 1$ and above $z=0$ only. The curves are means over eleven different time instants with a cadence of one minute and the shaded areas correspond to the standard deviations obtained from the temporal variations. Data points with $\lambda < 2.09\,\times\,10^{-2}\,\mathrm{Hz}$ are excluded. (\emph{b}) Mean magnetic field strength $\langle|\boldsymbol{B}|\rangle$ (red) and mean swirling strength $\langle\lambda\rangle$. The dark blue and red curves refer to the same MHD time instants as in (\emph{a}), while the light blue curve refers to $14$ time instants of a purely hydrodynamical simulation.
        }
        \label{fig:Correlation_Spearman}
    \end{figure*}

\section{Results and discussion}\label{sec:results}
       
    This section starts by taking a global view on the simulation, evidencing the correlation between vortices and magnetic field. It then concentrates on a single swirl event, which is shown to be a propagating torsional Alfv\'en pulse, and concludes with an estimate of the energy carried by the pulse and by swirling motions in the entire simulation domain. Both global and local analyses are supplemented by appendices.
    
    %
    %%%%%%%%%%%%%%%%%%%%%%%%%%%%%%%%%%%%%%%%%%%%%%%%%%%%%%%%%%%%%%%%%%%%%%%%%%%%%%
    %%%%%%%%%%%%%%%%%%%%%%%%%%%%%%%%%%%%%%%%%%%%%%%%%%%%%%%%%%%%%%%%%%%%%%%%%%%%%%
    %
    
    \subsection{Relation between swirls and magnetic fields} \label{subsec:DistSSinSim}

    Figure\,\ref{fig:SwirlsDistribution_TauBeta} shows a volume rendering of the distribution of swirling strength seen across a vertical section through the simulation domain at an arbitrary location and time instant. Three regions can be distinguished. The first one extends from the bottom of the box to $z\approx\!0\,\mathrm{km}$, the height that roughly corresponds to the surface of optical depth $\tau_{500}=1$ indicated in red color. This region coincides with the convection zone and it is characterized by an almost isotropic distribution of swirls, mostly produced by the convective and turbulent motions of the plasma. 
        
    At the photospheric and chromospheric level, there exist two distinct regions: The first one shows tubes of intense swirling strength aligned with the magnetic field, while the other one is characterized by a severe depletion of swirls. The yellow, funnel-shaped surface, which sharply separates these two regions, is given by the isosurface of plasma-$\beta = 1$, where $\beta = p_{\rm g}/p_{\rm m}$ is the ratio between gas pressure $p_{\rm g}$ and magnetic pressure $p_{\rm m}$. This clear separation shows evidence of strong coupling between swirling motions and magnetic fields. Above the yellow surface and within the funnels $\beta < 1$, therefore, the dynamics are governed by the magnetic fields, while the region of depleted swirling strength is characterized by $\beta > 1$, meaning that the gas pressure dominates. 
        
    We notice, however, that the region in between the magnetic flux concentrations in the photosphere is not completely void of swirls. There exist arches of swirling strength that are low lying and arches that reach farther out, as well as vertically directed swirls, which rise above the solar surface as it can be seen for example in the ranges $3000\,\mathrm{km} < x < 5000\,\mathrm{km}$ and $7000\,\mathrm{km} < x < 9000\,\mathrm{km}$ in Fig.\,\ref{fig:SwirlsDistribution_TauBeta}. This kind of swirls and swirl arches were also noted and described by
    \citet{2010NewA...15..460M} and % Muthsam et al. (2010)
    \citet{2011A&A...533A.126M} % Moll et al. (2011)
    and can probably to a great extent be identified with horizontal vortex tubes that form at the edges of granules
    \citep{2010ApJ...723L.180S}. % Steiner et al. (2010)
        
    Appendix\,\ref{app:comparison} compares the time instant of Fig.\,\ref{fig:SwirlsDistribution_TauBeta} with a magnetic field-free simulation and with a simulation of initial magnetic field strength of $200\,{\rm G}$. It also quantifies the orientation of swirls as a function of height.

    \begin{sidewaysfigure*}[p]
        \centering
        \includegraphics[width=23cm]{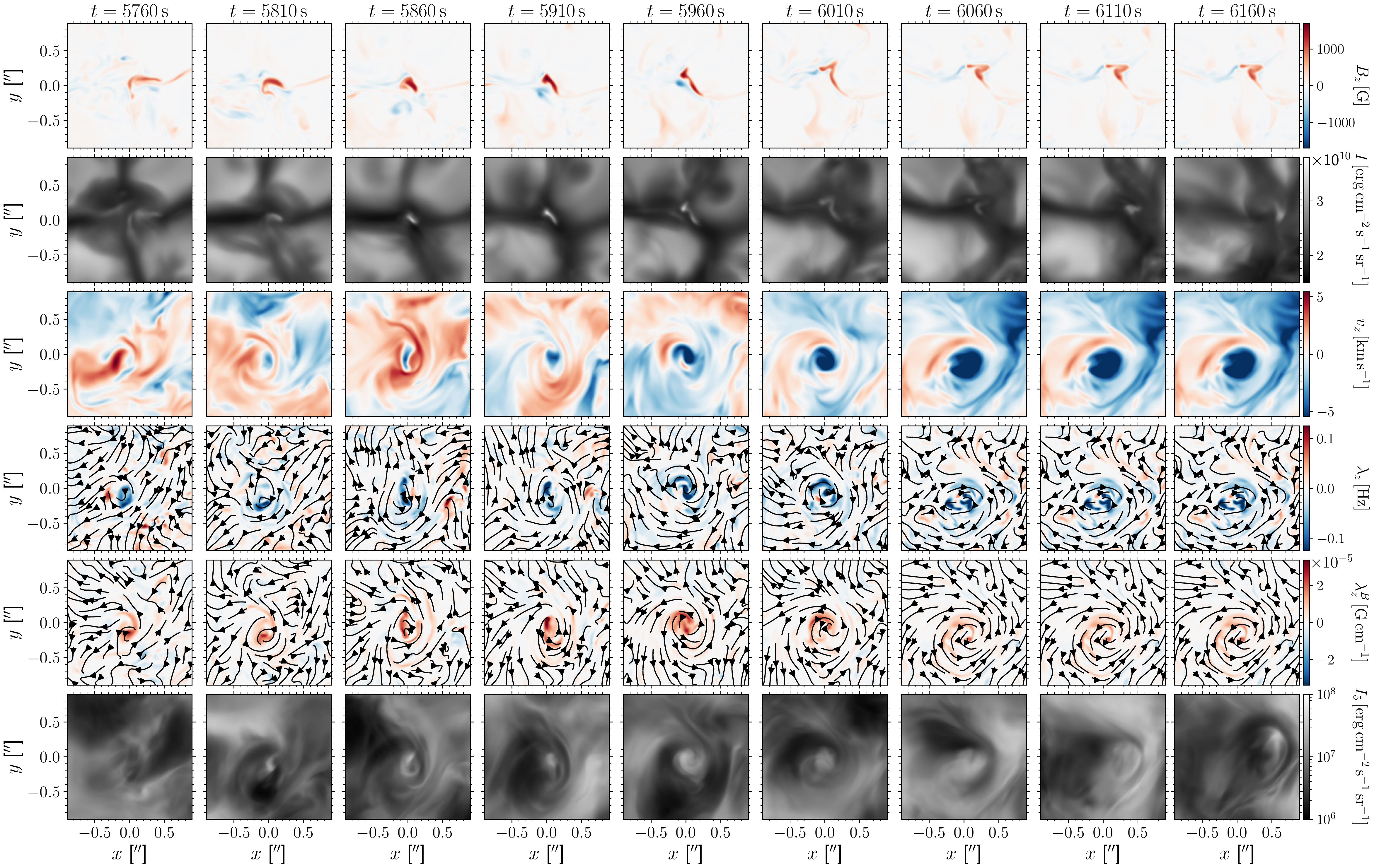}
        \caption{
        Time sequence of a single swirl event from $t=5760\,\mathrm{s}$ to $t=6160\,\mathrm{s}$. \emph{From top to bottom row}: Vertical component of the magnetic field $B_z$ at $z=0\,\mathrm{km}$, continuum intensity $I$, vertical velocity $v_z$ at $z=700\,\mathrm{km}$, vertical component of the swirling vector $\lambda_z$ at $z=700\,\mathrm{km}$, vertical component of the magnetic swirling vector $\lambda^{B}_z$ at $z=700\,\mathrm{km}$, and the bin-5 intensity $I_5$. Maps of $\lambda_z$ and $\lambda_z^{B}$ also show the streamlines of the velocity field and the magnetic field projected into the horizontal plane at $z=700\,\mathrm{km}$, respectively.
        }
        \label{fig:SingleSwirlEvent}
    \end{sidewaysfigure*} 

    Next, a statistical study is performed, with the purpose of evaluating the degree of correlation between the swirling strength $\lambda$, the magnetic field strength $|\boldsymbol{B}|$, and the magnetic swirling strength $\lambda^{B}$. 
    To do so, Spearman's $r$ rank coefficient is employed, which assesses how well two quantities are monotonically correlated: $r=1$ indicates perfect rank correlation, $r=-1$ stands for rank anti-correlation, while $r=0$ means that the two quantities are uncorrelated. Figure\,\ref{fig:Correlation_Spearman}a shows this coefficient, as a function of height $z$, for the correlation between $\lambda$ and $|\boldsymbol{B}|$, $r(\lambda,|\boldsymbol{B}|)$, and between $\lambda$ and $\lambda^{B}$, $r(\lambda,\lambda^{B})$. 

    From Fig.\,\ref{fig:Correlation_Spearman}a, we see that a strong rank correlation exists between swirling strength and magnetic swirling strength, $r(\lambda,\lambda^{B})$, in particular in the near surface convection zone and in the photosphere ($r \approx\!0.8$). Into the chromosphere, the correlation coefficient decreases to $r\approx\!0.6$. Therefore, we conclude that a strong correlation exists between $\lambda$ and $\lambda^{B}$ in the solar atmosphere and, consequently, also between plasma vortices and torsional perturbations in the magnetic field. The lower values in the chromosphere can be explained by the upper boundary condition of vanishing horizontal magnetic field component, which prevents torsional perturbations at the very top of the computational domain. 
    Given Eq.\,(\ref{eq:relationlambdas}), one can tentatively link this high correlation to torsional Alfvén waves, which is the topic of Sect.\,\ref{subsec:alfvén_pulses}. 

    Concerning the correlation between swirling strength and magnetic field strength, $r(\lambda,|\boldsymbol{B}|)$, Spearman's coefficient is $r\approx\!0.35$ in the convection zone, drops to $r\approx\!0.3$ in the photosphere and rises again to $r\approx\!0.4$ around the classical temperature minimum near $z=500$\,km before decreasing to $r\approx\!0.0$ in the middle chromosphere. The behavior of the curve is similar to $r(\lambda,\lambda^{B})$, but the overall value is always smaller and the variation is more pronounced. The low value in the convection zone can be explained by the turbulent motion of the plasma. Indeed, the source of vortices is not related to the strength of the magnetic field, which is almost everywhere much smaller than the kinetic equipartition value in the convection zone. Swirls that are carried from the convection zone to the photosphere rapidly lose angular velocity owing to the steep decrease in mass  density and correspondingly strong expansion of the plasma \citep{1997A&A...328..229N}. %Nordlund et al. (1997)
    This explains the rapid decrease in the average swirling strength above $z=0$\,km shown in Fig.\,\ref{fig:Correlation_Spearman}b. Together with the many nonmagnetic or weakly magnetized swirl arches in between the magnetic flux concentrations, this reduces the correlation between swirling strength and magnetic field in the photosphere. 
        
    However, the correlation stays high for swirls within the magnetic funnels ($\beta < 1$), which is demonstrated by the green curve in Fig.\,\ref{fig:Correlation_Spearman}a. While weak-field swirl-arches become less abundant with height, the correlation rises throughout the photosphere, reaching a maximum of $0.43$ at $z=500$\,km. Higher up in the atmosphere, the correlation $r(\lambda,|\boldsymbol{B}|)$ rapidly drops because the magnetic field becomes increasingly homogeneous and its strength drops, while the swirling strength steeply increases and remains highly structured as it can be seen from Fig.\,\ref{fig:Correlation_Spearman}b and Fig.\,\ref{fig:SwirlsDistribution_TauBeta}.\footnote{Despite the close to homogeneous magnetic field in the upper layers of the simulation domain, the swirling strength remains highly structured because the magnetic field dominates the dynamics everywhere and can produce and host swirls on small-scales; a finding recently also established by \citet{2020ApJ...894L..17Y} with MURaM simulations and by \citet[][Fig.\,5]{2020A&A...639A.118C} with CO$5$BOLD simulations.} This behavior is distinctly different from a magnetic field-free simulation in which the swirling strength does not show this steep increase (light blue curve in Fig.\,\ref{fig:Correlation_Spearman}b).
        
    %
    %%%%%%%%%%%%%%%%%%%%%%%%%%%%%%%%%%%%%%%%%%%%%%%%%%%%%%%%%%%%%%%%%%%%%%%%%%%%%%
    %
    
    \subsection{Evidence of Alfvén pulses}\label{subsec:alfvén_pulses}
       
    \begin{figure*}
        \resizebox{\hsize}{!}
        {\includegraphics{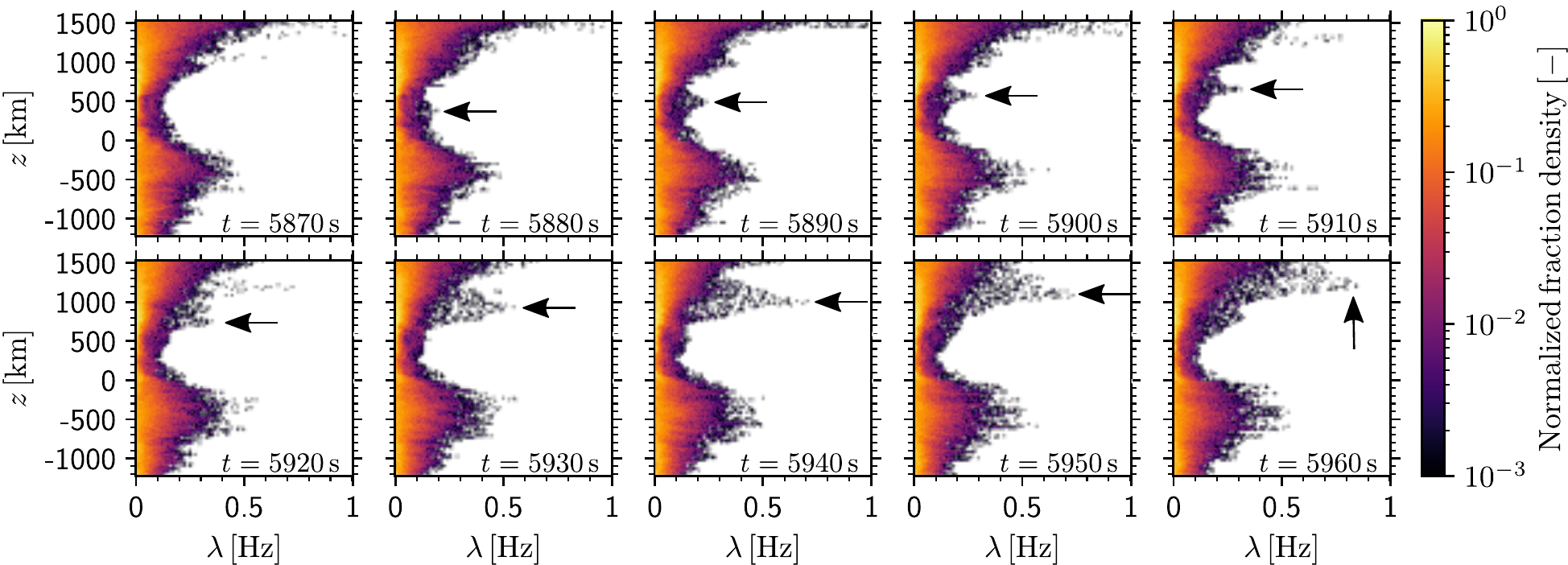}}
        \caption{
        Time sequence of the swirling strength distribution. Each panel represents a bi-dimensional histogram normalized to the maximum density fraction at each height level $z$. The bin sizes are $\Delta z = 10$\,km and $\Delta\lambda = 3.62\,\times10^{-3}\,{\rm Hz}$. These histograms refer to stacks of quadratic, plane-parallel cross sections of $1000\,\mathrm{km}$ side length centered
        on the swirl event of Fig.\,\ref{fig:SingleSwirlEvent}. Data points with $\lambda < 2.09\,\times\,10^{-2}\,\mathrm{Hz}$ are excluded. The black arrows point to the upward propagation of a local peak of swirling strength.
        }
        \label{fig:SSDIstrUpwardPropagation}
    \end{figure*}
    \begin{figure}
        \resizebox{\hsize}{!}{
            \includegraphics[width=\hsize]{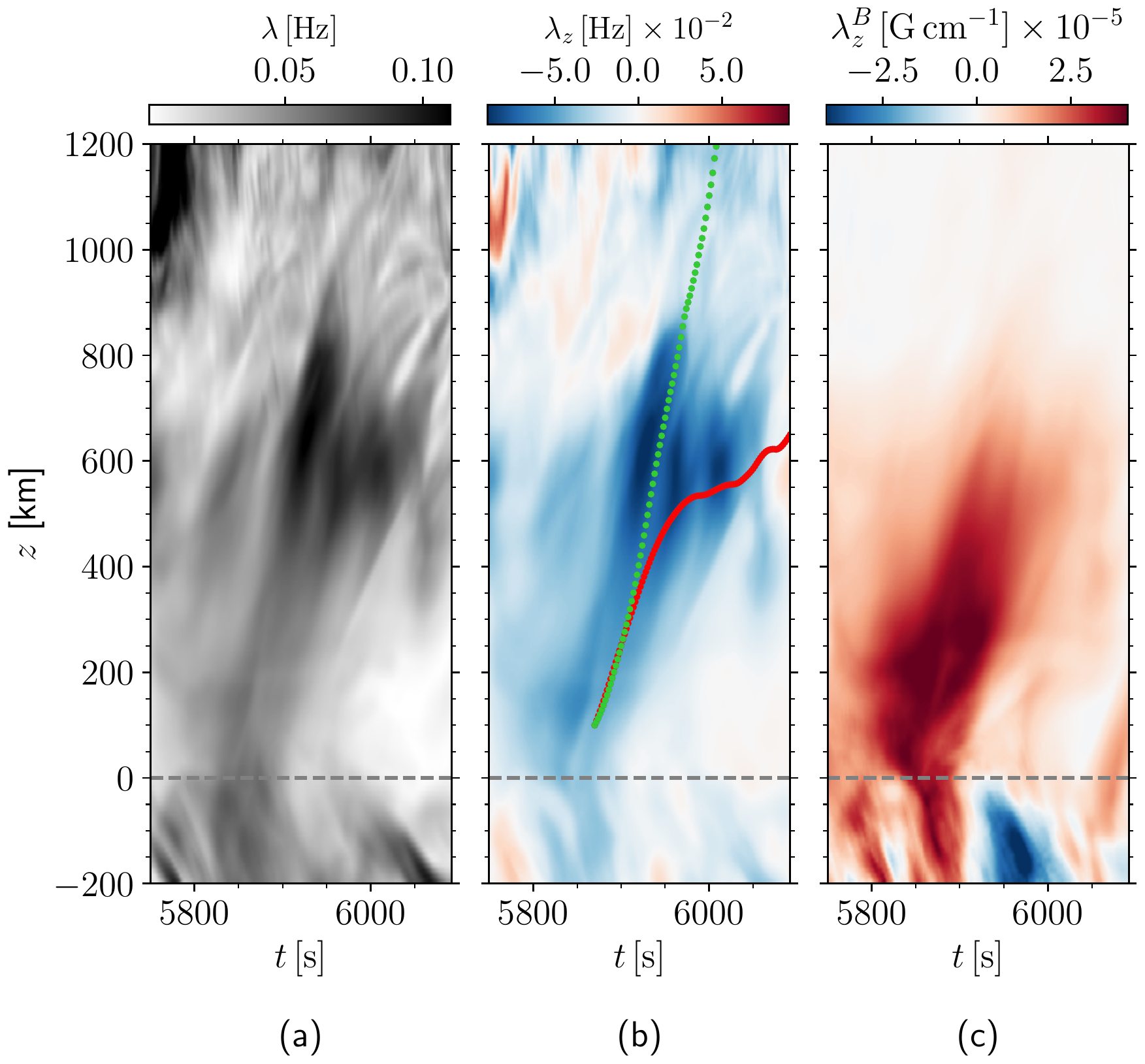}
        }
        \caption{
        Time-distance diagrams of (\emph{a}) the swirling strength $\lambda$, (\emph{b}) the vertical component of the swirling vector $\lambda_z$, and (\emph{c}) the vertical component of the magnetic swirling vector $\lambda^{B}_z$ for the swirl event shown in Fig.\,\ref{fig:SingleSwirlEvent}. Values at each time step and height are averaged over a finite horizontal plane of $150\,\mathrm{km}$ side length. The paths of two test particles moving with Alfv\'en plus bulk speed (green) and with sound plus bulk speed (red) along the swirl are shown in panel (b). 
        }
        \label{fig:SingleSwirlEvent_TDdiagrams}
    \end{figure}
    In this subsection, we study in detail one of the swirl events that occurred in the course of the simulation with the $50\,{\rm G}$ initial field strength. A list of eight supplementary events is presented in Appendix\,\ref{app:supplementary_swirls}. The six rows of Fig.\,\ref{fig:SingleSwirlEvent} show close-ups on various observable quantities that characterize the event. The cadence of the close-ups is $50\,\mathrm{s}$. The top row corresponds to a proxy of a magnetogram. It exhibits a positive polarity (upwardly pointing) magnetic flux concentration with $B_z\approx\!1500$\,G at $z=0\,\mathrm{km}$ and around $t = 5910\,{\rm s}$, while it has merely half that strength before and after this maximum. The flux concentration is responsible for the bright knot located within the integranular vertex visible in the continuum-intensity maps of the second row. Next to this flux concentration is a weaker one of inverse polarity. 
    
    The third, fourth, and fifth rows show, respectively, the vertical component of the velocity field, $v_z$, the vertical component of the swirling vector, $\lambda_z$, and the vertical component of the magnetic swirling vector, $\lambda_z^{B}$, all in a cross section at a height of $z = 700\,\mathrm{km}$. We see the development of a clockwise rotating swirl (negative $\lambda_z$) from the start of the time series, best visible from $t \approx\!5860\,\mathrm{s}$ onward in both $v_z$ and $\lambda_z$ (third and fourth rows). The positive polarity magnetic flux concentration in the top row and the BP in the second row appear to be rotating clockwise too, that is, in the same direction as the plasma does. The rotation is unidirectional (clockwise) over the full time period from $5760\,\rm{s}$ to $6160\,\rm{s}$, that is, over $400\,\rm{s}$. The fifth row exhibits that there also exists an azimuthal perturbation in the magnetic field but opposite to the velocity field. This aspect will be discussed in the paragraph after next. 
        
    Finally, the last row shows the  maps of the synthetic bin-5 intensity, $I_5$. It corresponds to the fifth opacity band of the non-gray radiative transfer used in the simulation and represents an average of intensities from strong spectral lines of large opacities \citep{Ludwig1992}. The $\tau_{500} = 1$ level computed with this opacity bin is located in the upper photosphere to lower chromosphere, and, therefore, $I_5$ can be taken as a proxy for chromospheric line core intensities.
    In these maps, we recognize a chromospheric swirl with a maximal diameter of $\approx\! 1.4\,\mathrm{arcsec}$, which is in the range of sizes of the ones reported by \citet{2009A&A...507L...9W}. % Wedemeyer-Bohm & Rouppe van der Voort

    From Fig.\,\ref{fig:SingleSwirlEvent} we can deduce an important aspect of this event: The vertical component of the swirling vector $\lambda_z$ and of the magnetic swirling vector $\lambda_z^{B}$ have opposite signs, that is, opposite orientations. In fact, it can be seen from the streamlines in the corresponding panels that the plasma is rotating in a clockwise fashion (negative swirling strength), while the magnetic field lines are counterclockwise twisted (positive magnetic swirling strength). This is a characteristic of upwardly propagating torsional Alfvén waves in a positive polarity (upwardly directed) magnetic field, as was pointed out in Sect.\,\ref{sec:theory}. Therefore, the fourth and fifth rows of Fig.\,\ref{fig:SingleSwirlEvent} represent the first piece of evidence of a perturbation similar to a torsional Alfvén wave in this swirl event.

    The time sequence of Fig.\,\ref{fig:SingleSwirlEvent} shows the evolution of the swirl in a horizontal section at $z=700\,\mathrm{km}$ only. In order to gain a perspective of its evolution in the vertical direction, Fig.\,\ref{fig:SSDIstrUpwardPropagation} gives the temporal evolution of the histograms of the swirling strength as a function of $z$. The histograms refer to a region of interest of $1.0 \times 1.0\,\mathrm{Mm}^2$ centered on the swirl and are taken with a cadence of $10\,\mathrm{s}$. This region of interest is almost as large as the individual maps of Fig.\,\ref{fig:SingleSwirlEvent}: It is large enough to ensure that the swirl stays within this region at all heights $0 < z < 1500\,{\rm km}$ and small enough to avoid disturbances from neighboring swirling motions. 
        
    From Fig.\,\ref{fig:SSDIstrUpwardPropagation} one recognizes a decline of swirls (both in density and in strength) at photospheric levels due to reasons already mentioned in Sect.\,\ref{subsec:DistSSinSim}. 
    However, there is a small bump of large swirling strengths appearing at $(t,z) = (5880\,\mathrm{s}, 400\,\mathrm{km})$, marked with an arrow. It grows and moves upward in time, proceeding $\approx\! 100\,\mathrm{km}$ every $10\,\mathrm{s}$. The bump that started in the photosphere then seems to merge with other local over-densities at $t=5930\,\mathrm{s}$ to form a large peak around $z=1000\,\mathrm{km}$, which continues its ascent after $t=5940\,\mathrm{s}$. We interpret the ascension of this over-density of large swirling strength to be the signature of an upwardly propagating swirl, which starts in the photosphere at $z \approx\!400\,\mathrm{km}$ and becomes manifest at chromospheric levels at around $t=5910\,\mathrm{s}$, as was already visible in Fig.\,\ref{fig:SingleSwirlEvent}. 
 
    To further investigate the upward propagation of the swirling motion, Fig.\,\ref{fig:SingleSwirlEvent_TDdiagrams} presents three time-distance diagrams. In these diagrams, the quantities at each height and instant (time-distance point) are averages over a horizontal plane cross-section of side length $150\,\mathrm{km}$ centered on the magnetic swirl event shown in Fig.\,\ref{fig:SingleSwirlEvent}. This side length corresponds to approximately $0.2\,{\rm arcsec}$, which is distinctly smaller than the side length of the maps shown in Fig.\,\ref{fig:SingleSwirlEvent}. Panel (a) shows the evolution of the swirling strength $\lambda$ and (b) the vertical component of the swirling vector $\lambda_z$. 
        
    One easily recognizes a crest of large swirling strength in both panels, starting at $(t,z)=(5830\,{\rm s}, 0\,{\rm km})$ and continuing up to at least $(t,z)=(5970\,{\rm s}, 1050\,{\rm km})$. The similarity between these two diagrams tells us that the swirl axis is to a large degree vertically directed. The crest in $\lambda_z$ fades above $z\approx\! 900\,{\rm km}$, while that of $\lambda$ can be followed further up to $z\approx\! 950\,{\rm km}$.
    A more careful analysis reveals that this disappearance is not due to a fading of the swirl itself but to the bending of the swirl out of the strictly vertical detection slit of $150\times150\,\mathrm{km}^2$ cross section used to establish these time-distance plots. In fact, from 
    Fig.\,\ref{fig:SSDIstrUpwardPropagation}, which is based on a wider cross-section, it can be readily seen that the swirl propagates up to the top of the computational domain, showing still a peak in swirling strength at $(t,z)=(5960\,{\rm s}, 1200\,{\rm km})$.

    From panels (a) and (b) of Fig.\,\ref{fig:SingleSwirlEvent_TDdiagrams} it can firstly be seen that this swirl is not a rigidly rotating magnetic structure but it propagates wave-like throughout the atmosphere in the upward direction. However, at any time, there is swirling motion distributed over a large height range and only the local maximum (the crest) of swirling strength is confined to a narrower height range. Second, we notice from panel (b) that the swirling strength is always negative, thus the rotation of the swirl is clockwise. Only in chromospheric layers above $800\,{\rm km}$ there are traces of counterclockwise motion, but these do not  arise from a precursory wave train starting from the photosphere. Thus, we do not deal with an oscillatory wave but with an unidirectional pulse. In fact, there exists a sequence of equally directed pulses as it can be best seen from panel (b) at around $z=600\,{\rm km}$ with a sequence of local maxima in swirling strength at times $t=5920\,{\rm s}$, $5970\,{\rm s}$, and $6020\,{\rm s}$.
        
    Panel (c) of Fig.\,\ref{fig:SingleSwirlEvent_TDdiagrams} shows the evolution of the vertical component of the magnetic swirling vector $\lambda_z^{B}$. From it, it is possible to observe that the propagation of negative $\lambda_z$ occurs in parallel with the propagation of positive $\lambda_z^{B}$. This behavior confirms that the vortex flow in the plasma is linked to a twist of opposite orientation in the magnetic field, which is a characteristic of torsional Alfvén waves propagating in direction of the magnetic field, as was already pointed out in the discussion of Fig.\,\ref{fig:SingleSwirlEvent}. 
        
    The value of $\lambda_z^{B}$ is large in the photosphere and then fades away in the chromosphere for three reasons. First, the magnetic field is weaker and therefore the value of $\lambda_z^{B}$ is smaller higher up in the atmosphere than in the photosphere. Second, the magnetic swirl becomes curved in the chromosphere bending outside of the detection slit used for establishing this time-distance diagram. Finally, the boundary conditions impose the magnetic field to be vertically directed at the top boundary, hence twists in the upper layers are suppressed. We also note that the twist angle of the magnetic field (inclination with respect to the vertical direction) does not have to be large but can be imperceptibly small where $\beta \ll 1$.
        
    Next, the propagation speed of the swirl is investigated. The green dotted curve of Fig.\,\ref{fig:SingleSwirlEvent_TDdiagrams}b shows the path of an imaginary test particle moving at speed $v(z) = \langle v_{\rm A}(z) \rangle \cdot \cos{(\langle \theta \rangle)} + \langle v_z(z) \rangle$, that is, with the mean local Alfvén speed plus the mean local, vertical plasma bulk-speed, starting in the photosphere at $z\!\approx\! 100\,{\rm km}$ from the ridge that leads highest up. Here, the mean is taken over the horizontal plane cross-section of side length $150\,\mathrm{km}$ of the detection slit and $\langle\theta\rangle$ is the mean inclination of the swirl, that is, the angle of the swirling strength vector with respect to the $z$-axis. The red particle moves with speed $v(z) = \langle c_{\rm s}(z)  \rangle  \cdot \cos{(\langle \theta \rangle)} + \langle v_z(z) \rangle$, which corresponds to the mean sound plus plasma bulk-speed. It starts at the same position as the green curve.
        
    We notice how the green trajectory perfectly follows the propagation of the vertical component of the swirling vector, which means that the swirl is moving with the local Alfvén plus bulk speed. Different from that, the red curve strongly deviates from the ridge of maximal vertical swirling strength starting from approximately $z=400\,{\rm km}$, indicating that the swirl does not propagate with sound speed. This is the second piece of evidence of the Alfvénic nature of the swirl event.

    It is essential to notice that the vertical component of both the swirling vector and the magnetic swirling vector do not change orientation in time. This means that the swirling motion, as well as the perturbation of the magnetic field, are unidirectional. Therefore, one must think of a torsional Alfvén pulse instead of a generic, oscillatory, torsional Alfvén wave. The existence of such torsional pulses in the solar atmosphere has been suggested by \citet{2019NatCo..10.3504L} % Liu et al, (2020)
    using numerical simulations with a torsional perturbative driver acting on a preexisting magnetic flux tube. Here, it has been demonstrated that such torsional Alfvén pulses also emerge from a fully self-consistent, realistic, numerical simulation of the solar atmosphere. 
    
    The present section focused on a single, isolated swirl event. More often, simulations show swirling motions in the chromosphere to appear as a result of the superposition of two or more individual swirls. Examples of the superposition of swirls are displayed and discussed in Appendices\,\ref{app:superposition_of_swirls} and \ref{app:supplementary_swirls}.
    
    %
    %%%%%%%%%%%%%%%%%%%%%%%%%%%%%%%%%%%%%%%%%%%%%%%%%%%%%%%%%%%%%%%%%%%%%%%%%%%%%%
    %

    \subsection{Dynamics of the pulse}
        
    \begin{figure*}[h!]
        \resizebox{\hsize}{!}
        {\includegraphics{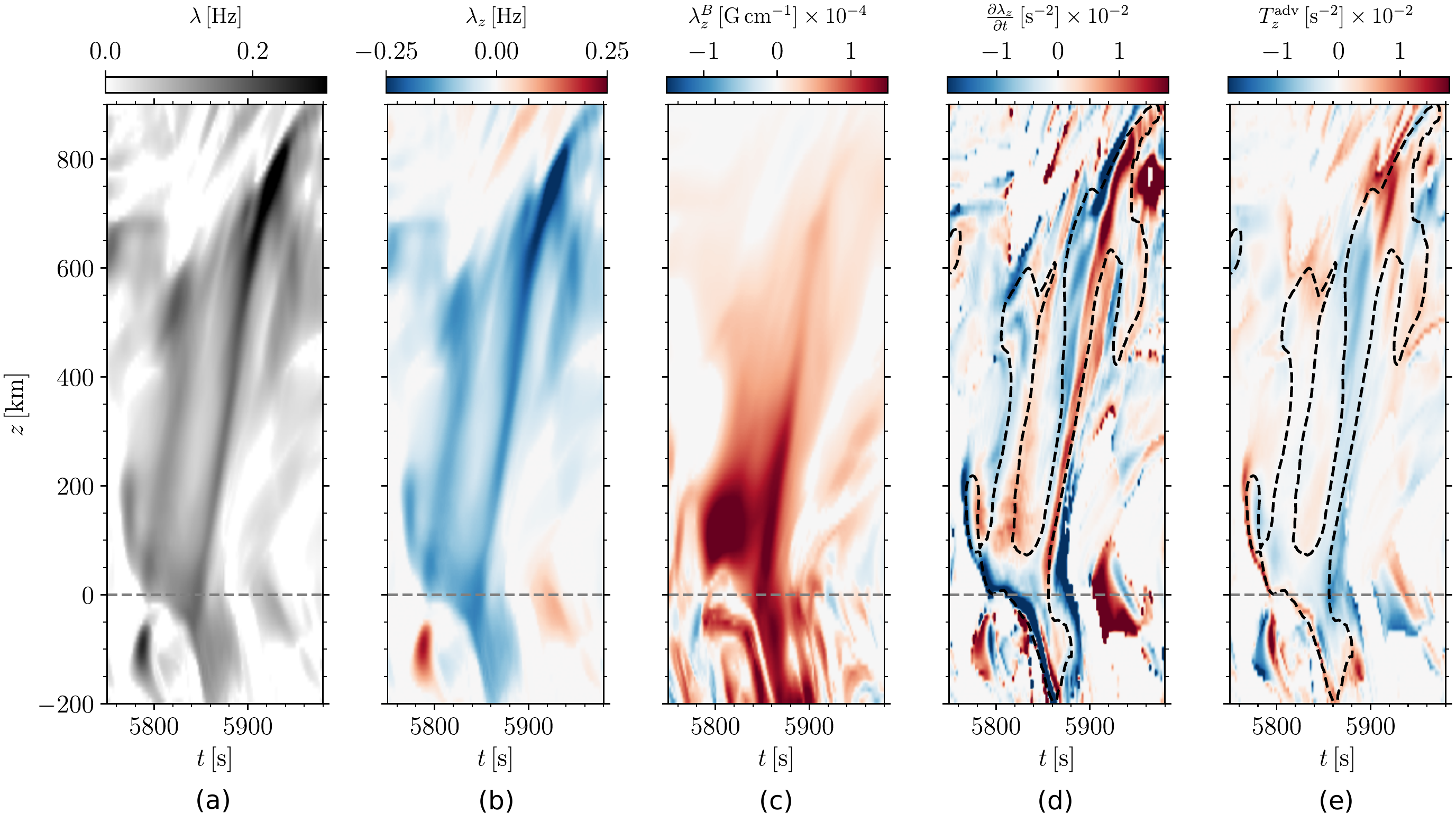}} \\[1em]
        \resizebox{\hsize}{!}
        {\includegraphics{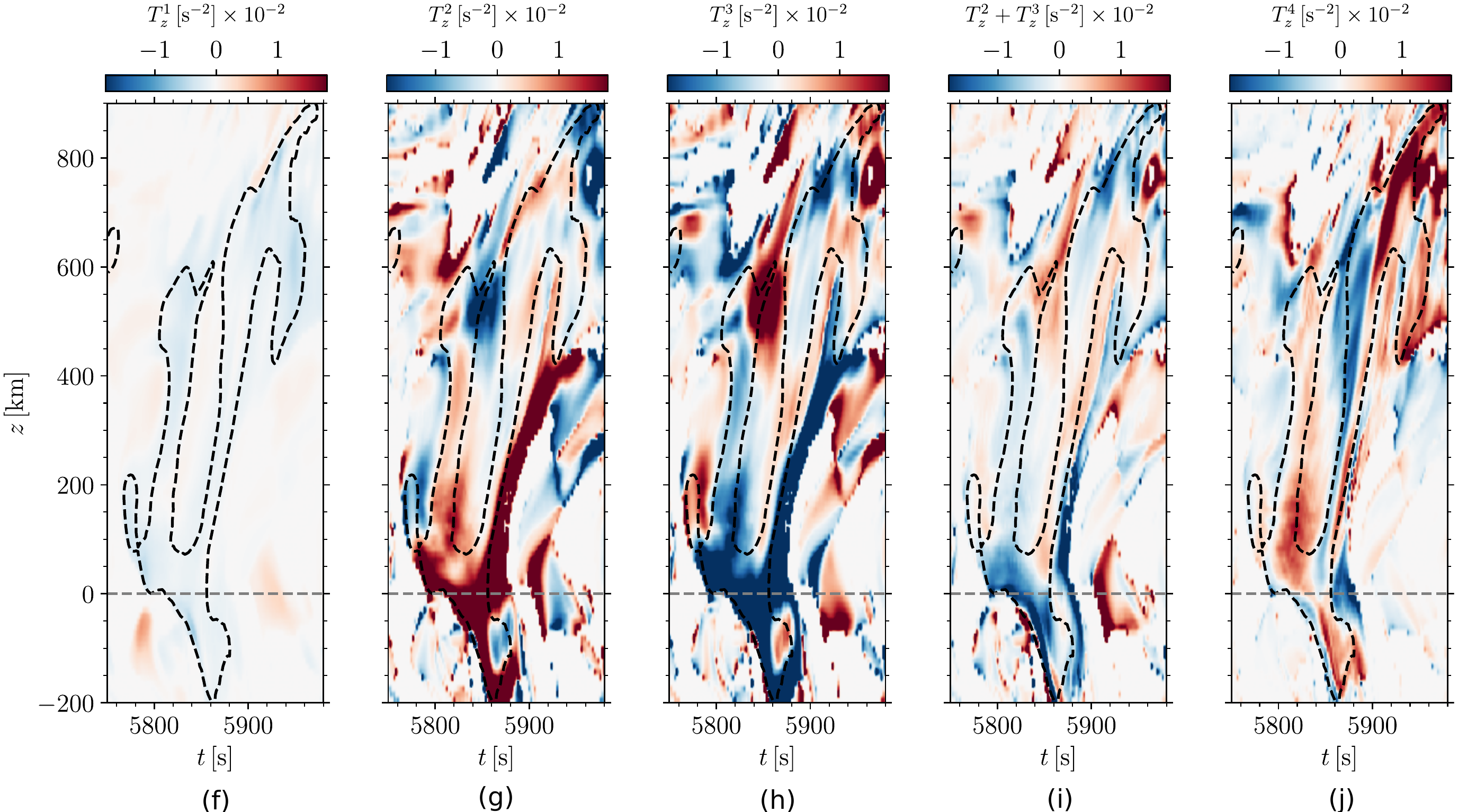}}
        \caption{
        Time-distance diagram of (\emph{a}) the swirling strength $\lambda$ and of the vertical component of the following quantities: (\emph{b}) swirling vector $\lambda_z$, (\emph{c}) magnetic swirling vector $\lambda_z^{B}$, (\emph{d}) partial derivative of the swirling vector ${\rm \partial}\lambda_z/{\rm \partial}t$, (\emph{e}) advection term $T_z^{\rm adv}=[(\boldsymbol{v}\cdot\nabla)\boldsymbol{\lambda}]_z$, (\emph{f}) stretching term $T_z^1$, (\emph{g}) hydrodynamical baroclinic term $T_z^2$, (\emph{h}) magnetic baroclinic term $T_z^3$, (\emph{i}) sum of baroclinic terms $T_z^{\rm bar} = T_z^2 + T_z^3$, and (\emph{j}) magnetic tension term $T_z^4$. The black dashed contours correspond to the $\lambda_z = -0.1\,{\rm Hz}$ contours, which highlight the propagation of the Alfvén pulse. All plots are relative to the swirl event shown in Fig.\,\ref{fig:SingleSwirlEvent} and are derived from a fixed, one pixel wide vertical slit through the center of that swirl. 
        }
        \label{fig:SingleSwirlEvent_TDdiagrams_SingleLOS}
    \end{figure*}
    This section focuses on the dynamics of the magnetic swirl shown in Fig.\,\ref{fig:SingleSwirlEvent}, in particular, on the physical processes responsible for the production and propagation of the swirling motion and Alfvén pulse. For this purpose, we employ the swirling equation, Eq.\,(\ref{eq:swirling_eq}), to reveal the local production of the vertical component of the swirling vector by the various source terms. Each one of these source terms is related to a different physical mechanism, as described in Sect.\,\ref{subsec:vortex_detection}. 
    
    Figure\,\ref{fig:SingleSwirlEvent_TDdiagrams_SingleLOS} shows time-distance diagrams of various quantities. Different from Fig.\,\ref{fig:SingleSwirlEvent_TDdiagrams}, the time-distance slit in this case is given by a single, vertical column of computational cells (single line-of-sight), located in the center of the magnetic swirl. Panel (a) shows the time-distance plot for the swirling strength $\lambda$, (b) the vertical component of the swirling vector $\lambda_z$, and (c) the vertical component of the magnetic swirling vector $\lambda_z^{B}$, similar to Fig.\,\ref{fig:SingleSwirlEvent_TDdiagrams}. However, with the single line-of-sight slit, we see two separate pulses of swirling strength, both with the same sense of rotation. They propagate upward at local Alfvén speed: the first starting at $(t,z) = (5800\,{\rm s}, 0\,{\rm km})$, the second at $(t,z) = (5850\,{\rm s}, -50\,{\rm km})$. These pulses are well paired with perturbations in the magnetic field, as can be seen from panel (c). These plots reproduce in more detail what was already observed from Fig.\,\ref{fig:SingleSwirlEvent_TDdiagrams}.

    Figures \ref{fig:SingleSwirlEvent_TDdiagrams_SingleLOS}d and \ref{fig:SingleSwirlEvent_TDdiagrams_SingleLOS}e show the time-distance diagrams of the local production and the advection of vertical component of the swirling vector, $\partial \lambda_z / \partial t$ and $[(\boldsymbol{v}\cdot\nabla)\boldsymbol{\lambda}]_z$,  
    respectively. Together, these two quantities constitute the material derivative that appears on the left-hand side of Eq.\,(\ref{eq:swirling_eq}). We notice that the local production of negative swirling strength between $-200\,\mathrm{km} \lesssim z \lesssim 200\,\mathrm{km}$ and $5760\,\mathrm{s} \lesssim t \lesssim 5860\,\mathrm{s}$ in panel (d) corresponds to the appearance of negative swirling strength in panel (b). The black dashed lines are contours of $\lambda_z = -0.1\,{\rm Hz}$. They are intended to visualize the boundaries of the two pulses and guide the eye in identifying the relevant source  terms. Also from panel (d), we notice that the two upwardly propagating pulses are characterized by the production of negative swirling strength on the left (leading) boundary of the two stripes, which is compensated for, approximately $20\,{\rm s}$ later, by the production of positive swirling strength on the right (trailing) boundary. We conclude that these two sources cause the upward evolution of the swirl. 
        
    Contributions from the advection term to the local production of swirling strength are mostly subdominant in the photosphere and into the low chromosphere, as one can see from panel (e). Therefore, we conclude that the swirling strength is mainly locally produced in these regions. Only in the upper part of the chromospheric layers,  the advection term produces some non-negligible positive vertical swirling strength\footnote{To be precise, the advection term only advects swirling strength in or out of the time-distance slit; it does not produce it. Therefore, a positive value of $[(\boldsymbol{v}\cdot\nabla)\boldsymbol{\lambda}]_z$ can indicate either the advection of positive swirling strength into the slit, or negative swirling strength out of it.}. This production can only contribute to the reduction of the strong pulse in the height range $600\,{\rm km} < z < 800 \,{\rm km}$ around $t=5920\,{\rm s}$ and to the fading of the tip of swirling strength around $(t,z) = (5900\,{\rm s}, 800\,{\rm km})$, but it is not the main cause. 
    %Therefore, if the dynamics cannot explain the fading of the swirl, we conclude that its structure must be intrinsically tilted, as already hypothesized in Sect.\,\ref{subsec:alfvén_pulses}.
    Therefore, if the dynamics of the swirling strength alone cannot explain the fading of the swirl beyond 850\,km, we conclude that it must be the tilted swirl axis to cause it, as already hypothesized in Sect.\,\ref{subsec:alfvén_pulses}. This tilt also leads to the reduction of the vertical propagation speed of the swirl, seen to set in already beyond $z\approx 600$\,km.
        
    The second row of Fig.\,\ref{fig:SingleSwirlEvent_TDdiagrams_SingleLOS} shows the time-distance diagrams of the various terms of Eq.\,(\ref{eq:swirling_eq}). It serves to infer which physical processes are responsible for the production of negative swirling strength at the photospheric level and for the propagation of the pulses. 
    The gravitational term $T^5_z$ is omitted because the simulations employ a constant gravitational field and thus $T^5_z=0$.
    First of all, analyzing the different source terms, we notice that the stretching term, $T^1_z$, is much weaker than the other terms. Second, the two baroclinic terms, $T^2_z$ and $T^3_z$, have similar configurations but are of opposite sign. Therefore, panel (i) shows the sum $T^2_z + T^3_z$ from which it can be seen that most of their contributions cancel out. 
    This result is in accordance with the findings of \citet{2020A&A...639A.118C}. % Canivete Cuissa & Steiner, 2020
        
    The counter-balancing of the two baroclinic effects inside and across the boundary of a stationary magnetic flux tube can be explained by the required equilibrium between magnetic pressure, magnetic tension forces, and gas pressure. However, the magnetic flux tube hosting the swirl is not stationary as can be seen from Fig.\,\ref{fig:SingleSwirlEvent}. The dynamics of the magnetic structure and of the surrounding plasma can break the balance between the pressures, which can result in a net production of swirling strength, as visible from panel (i). Thus, it becomes clear that the local production of negative swirling strength seen in panel (d) between $-200\,\mathrm{km} \lesssim z \lesssim 100\,\mathrm{km}$ and $5780\,\mathrm{s} \lesssim t \lesssim 5860\,\mathrm{s}$ is caused by the imbalance between the two baroclinic terms. More specifically, around $z=0\,{\rm km}$ and between $5790\,{\rm s}\,\lesssim t \lesssim 5850\,{\rm s}$ the baroclinic magnetic term, $T^3_z$, which is producing negative swirling strength, overcomes the hydrodynamical one, $T^2_z$, which instead produces positive swirling strength. 
        
    Because the time-distance slit is located in the center of the magnetic flux concentration, where the $\beta = 1$ surface dips below $z=0\,{\rm km}$ into the convection zone (see Fig.\,\ref{fig:SwirlsDistribution_TauBeta}), the magnetic field dominates the gas pressure and the dynamics in this location. This suggests that the origin of the swirl is due to magnetic effects alone that could arise from an MHD instability or from magnetic annihilation, possibly owing to nearby inverse polarity fields such as in the case of quiet Sun Ellerman bombs \citep[QSEBs,][]{2020A&A...641L...5J}. % Joshi et al. (2020)
    Alternatively, the dominance of the baroclinic magnetic term may also arise from a weakening of the hydrodynamic baroclinic term caused by, for example, a sudden low pressure region or the bath tub effect \citep[see][]{2012Natur.486..505W} % Wedemeyer et al., 2014 
    in the convective motion surrounding the magnetic flux concentration. Revealing the true origin of the swirls calls for a multidimensional analysis of the different source terms of swirling strength, which goes beyond the scope of the present paper but shall become subject of a subsequent study.
        
    Having dealt with the triggers of such an event, we next pay attention to the propagation of the swirl. As already mentioned further above, the upward propagation is caused by the production of negative swirling strength and its successive destruction by the production of a positive counter-part, approximately 20\,s later. The inspection of the source-term diagrams of Fig.\,\ref{fig:SingleSwirlEvent_TDdiagrams_SingleLOS} reveals that the magnetic tension term, $T^4_z$, is the main responsible for the production and destruction of $\lambda_z$ along the two pulses of swirling strength, as can be seen from Fig.\,\ref{fig:SingleSwirlEvent_TDdiagrams_SingleLOS}j. The stretching term, $T^1_z$, plays no role in the propagation of the swirl, the hydrodynamical baroclinic term $T^2_z$ is overcome by its magnetic counterpart $T^3_z$, which in turn only partially contributes to the production of the first pulse of swirling strength and to the destruction of the second.
    Therefore, we can conclude that magnetic tension forces are responsible for the propagation of the swirl in the photosphere and low chromosphere. This analysis provides further evidence of the Alfvénic nature of the vortex flow since the magnetic tension is the driving force of Alfvén waves.

    %
    %%%%%%%%%%%%%%%%%%%%%%%%%%%%%%%%%%%%%%%%%%%%%%%%%%%%%%%%%%%%%%%%%%%%%%%%%%%%%%
    %

    \subsection{Energetic considerations}\label{subsec:energetics}
    \begin{figure}
        \resizebox{\hsize}{!}
        {\includegraphics{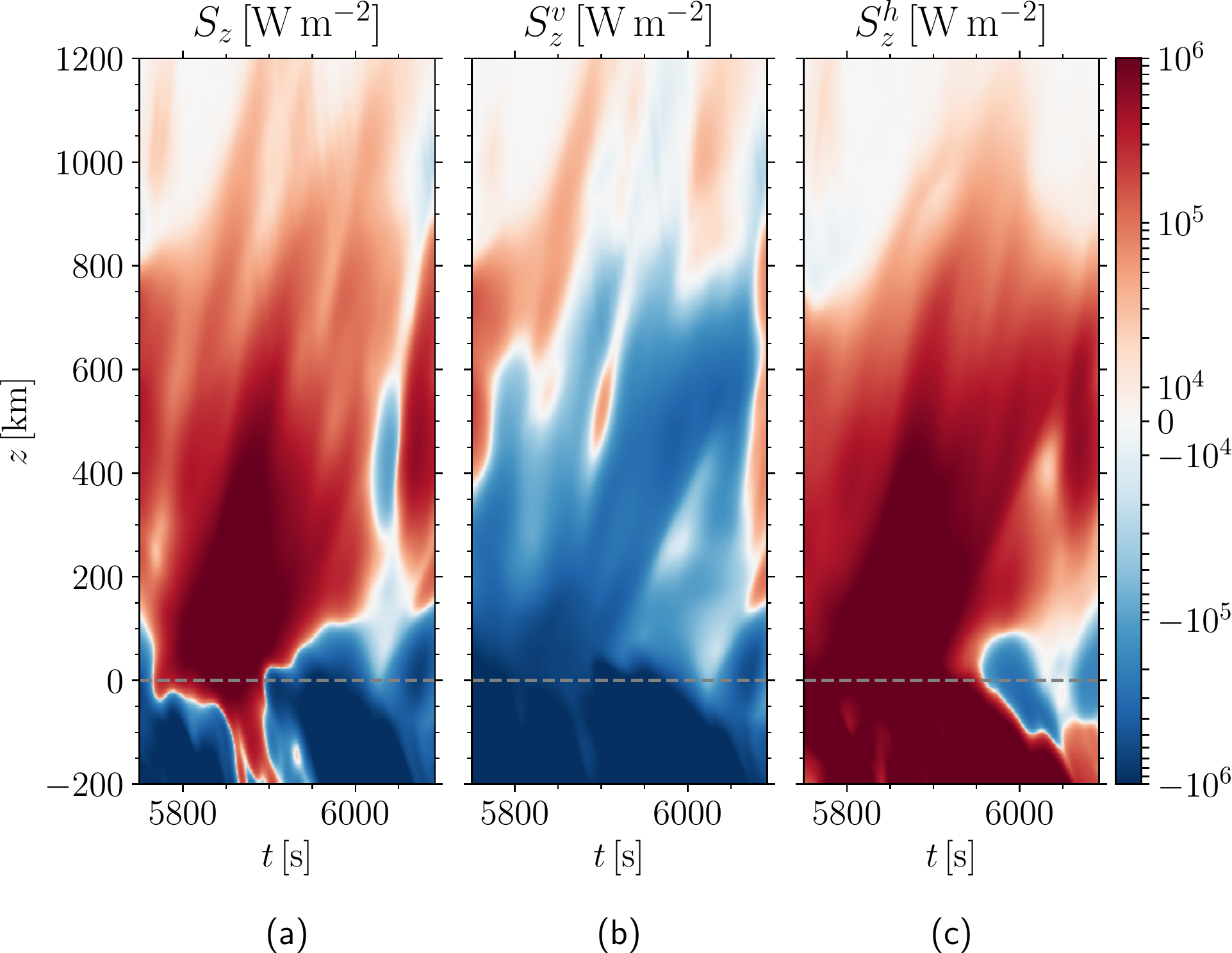}}
        \caption{
        Time-distance diagrams of the vertical component of the Poynting flux vector $S_z$ (\emph{a}) and of the two terms that compose it, $S_z^{\rm v}$ (\emph{b}) and $S_z^{\rm h}$ (\emph{c}), for the swirl event shown in Fig.\,\ref{fig:SingleSwirlEvent}. Values at each time step and height level are averages over a finite horizontal plane of $150\,\mathrm{km}$ side length. 
        }
        \label{fig:SingleSwirlEvent_TDdiagrams_Poynting}
    \end{figure}
    The purpose of the present section is to investigate how much energy swirling events and associated Alfvén pulses can channel up from the photosphere to the upper atmospheric layers. To this end, we analyze the vertical component of the Poynting flux vector and the mechanical energy flux relative to the swirling motions, as given in Sect.\,\ref{subsec:energyflux}. The first subsection concentrates on the single swirl event of Fig.\,\ref{fig:SingleSwirlEvent} and the second one evaluates the mean Poynting flux over the full computational domain.
    
    %
    %%%%%%%%%%%%%%%%%%%%%%%%%%%%%%%%%%%%%%%%%%%%%%%%%%%%%%%%%%%%%%%%%%%%%%%%%%%%%%
    %    
    
    \subsubsection{Energy transported by the swirl of Fig.\,\ref{fig:SingleSwirlEvent}}\label{susubbsec:energetics1}
    Figure\,\ref{fig:SingleSwirlEvent_TDdiagrams_Poynting}a shows the time-distance diagram
    of the vertical component of the Poynting flux, $S_z$, associated with the swirl event of  Fig.\,\ref{fig:SingleSwirlEvent}. Like in 
    Fig.\,\ref{fig:SingleSwirlEvent_TDdiagrams}, the distance slit stretches from $z = -200$\,km to $z = 1200$\,km. It has a finite horizontal, quadratic extension of $150\,\mathrm{km}$ side length, and $S_z$ is taken to be the average over this finite cross section. At the location of origin of the Alfv\'en pulse at around $z=0$ 
    (as was seen from Fig.\,\ref{fig:SingleSwirlEvent_TDdiagrams_SingleLOS}), the upwardly directed Poynting flux
    is on the order of $10^6\,\mathrm{W}\,\mathrm{m}^{-2}$.
    The flux decreases with height but it is still on the order of $10^5\,\mathrm{W}\,\mathrm{m}^{-2}$ at chromospheric levels. Panels (b) and (c) of Fig.\,\ref{fig:SingleSwirlEvent_TDdiagrams_Poynting} show the corresponding time-distance diagrams of the terms $S_z^{\rm v}$ and $S_z^{\rm h}$ respectively, which are defined in Eq.\,(\ref{eqn:z-Poynting}). We notice that the upward flux is entirely due to the term $S_z^{\rm h}$, which means that the energy is carried by the horizontal motions of the magnetized plasma. Hence, the magnetic swirl and associated Alfvén pulse are responsible for the upwardly directed energy transfer. 

    \begin{figure}
        \resizebox{\hsize}{!}
        {\includegraphics{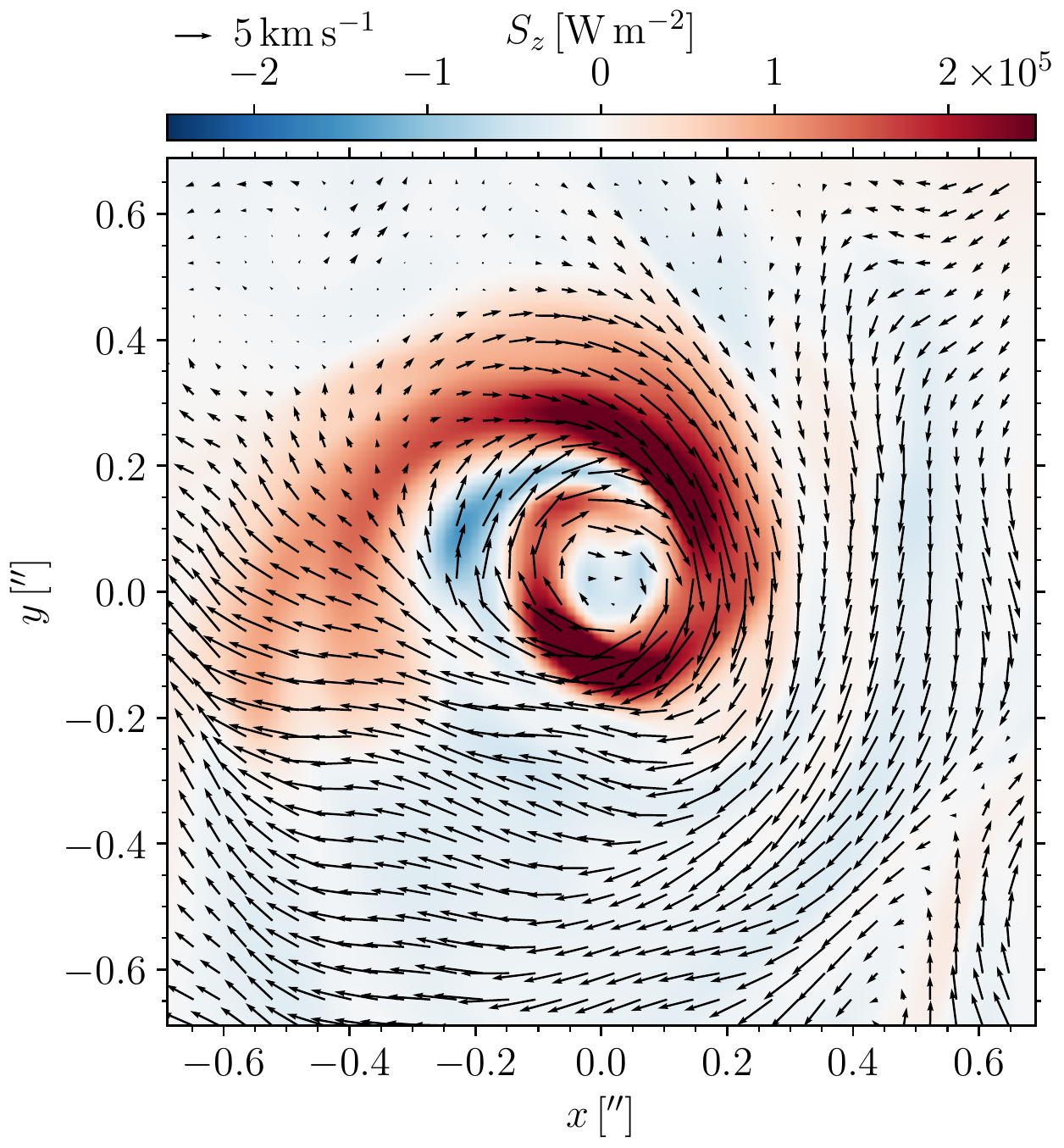}}
        \caption{
        Vertical component of the Poynting flux vector $S_z$ of the swirl event presented in Fig.\,\ref{fig:SingleSwirlEvent} in a horizontal section at $z=700\,\mathrm{km}$. Arrows indicate the velocity vector field projected into the horizontal plane. Their length is proportional to the magnitude of the horizontal flow.
        }
        \label{fig:SingleSwirl_HorizontalSection_Poynting}
    \end{figure}

    Figure \ref{fig:SingleSwirl_HorizontalSection_Poynting} reveals the spiral pattern outlined by the vertical component of the Poynting flux, $S_z$, in a horizontal section centered on the swirl and at $(t,z) = (5980\,{\rm s},\,700\,\mathrm{km})$. It roughly follows that of the swirling plasma flow, as it can be seen from the overlaid velocity field. This reinforces the idea that the swirl is carrying magnetic energy through the solar atmosphere. The maximum positive Poynting flux of the swirl is $S_{z,{\rm max}} = 3.5\times10^5\,\mathrm{W}\mathrm{m}^{-2}$, while the mean vertical Poynting flux is $S_z = 34.5\,\mathrm{kW}\mathrm{m}^{-2}$,
    where the mean was taken over a circular area of diameter $1.2\,{\rm arcsec}$
    centered on the swirl and over the time period from $t = 5760\,{\rm s}$ to $t = 6050\,{\rm s}$.

    From these numbers, the contribution of the swirl alone to the mean vertical Poynting flux through the cross section of the full computational domain of $9.6\times 9.6\,{\rm Mm}^2$ is $236\,{\rm W}\,{\rm m}^{-2}$. This value is smaller but on the order of the $440\,{\rm W}\,{\rm m}^{-2}$ reported by \citet{2012Natur.486..505W} % Wedemeyer-Bohm et al, 2012
    for the contribution to a cross section of $8\times 8$\,Mm$^2$, however, at a height of 2000\,km, that is,  at the base of the transition zone. 
    We attribute this weak Poynting flux of the swirl of Fig.\,\ref{fig:SingleSwirlEvent} to its relatively weak magnetic field and to its bending above $z\approx 700\,{\rm km}$. 
    
    \citet{2019NatCo..10.3504L} count from observations with the \ion{Ca}{ii} H Broadband Filter Imager (BFI) of the Solar Optical Telescope (SOT) of Hinode $\bar{N}_{\rm s} = 48.2$ swirls at any time in a field of view of $A = 800\,\mathrm{Mm}^2$. Assuming that these swirls are similar to the one of Fig.\,\ref{fig:SingleSwirlEvent} with a mean Poynting flux of $S_z = 34.5\,{\rm kW}\,{\rm m}^{-2}$ over a circular cross section of diameter $1.2\,{\rm arcsec}$, we obtain from Eq.\,(\ref{eq:averagesflux}) an average energy flux of
    $\bar{S}_z = 1.2\,{\rm kW}\,{\rm m}^{-2}$ at a height of $z=700\,{\rm km}$. This flux is not enough to compensate the radiative losses of approximately
    $4.3\,{\rm kW}\,{\rm m}^{-2}$ (excluding Lyman $\alpha$) in the
    semiempirical model of the chromosphere by \citet{1981ApJS...45..635V} %Vernazza et al (1981)
    but could still account for an important source of energy in the upper chromosphere and the corona.
        
    In the same way as for the Poynting flux, it is possible to estimate the mean mechanical energy flux owing to swirls. We take again the single swirl of Fig.\,\ref{fig:SingleSwirlEvent} as a model and average again over a fixed circular area of diameter $1.2\,{\rm arcsec}$ and the time period from $t=5760\,{\rm s}$ to $t=6050\,{\rm s}$. Furthermore, we assume as propagation speed the local Alfv\'en plus bulk speed in the vertical direction, $v_{\rm p} = v_{\rm A}\cos{(\theta)} + v_z$, as done in Sect.\,\ref{subsec:alfvén_pulses}, to compute the mechanical energy fluxes according to Eqs.\,(\ref{eq:mechflux_vh}) and (\ref{eq:mechflux_vlambda}).
        
    Using Eq.\,(\ref{eq:mechflux_vh}), the estimate for the mean  vertical mechanical flux owing to horizontal velocities is $F_z^{\rm r} = 6.1\,{\rm kW}\,{\rm m}^{-2}$, while Eq.\,(\ref{eq:mechflux_vlambda}) yields $F_z^{\rm \lambda} = 15.2\,{\rm kW}\,{\rm m}^{-2}$. 
    These formulas give only an approximate upper limit of the true mechanical energy flux because of a radial velocity component that was not removed in the computation of $F_z^{\rm r}$, while $F_z^{\lambda}$ is overestimated because of contribution from peripheral swirling strength not pertaining to the swirl under consideration. Nevertheless, the computed mechanical flux is only a fraction of the Poynting flux, which reflects the circumstances that the plasma-$\beta \lesssim 1$ in the considered region. 

    Assuming, once again, that the swirl shown in Fig.\,\ref{fig:SingleSwirlEvent} is representative of the $\bar{N}_{\rm s} = 48.2$ swirls detected at any time in a field-of-view of $A = 800\,{\rm Mm}^2$, and using Eq.\,(\ref{eq:averagemflux}) to compute the upper limit of the average mechanical energy flux owing to swirls, we find, at $z=700\,{\rm km}$, $\bar{F}_z^{\rm r} = 218\,{\rm W}\,{\rm m}^{-2}$ and $\bar{F}_z^{\rm \lambda} = 544\,{\rm W}\,{\rm m}^{-2}$, using Eq.\,(\ref{eq:mechflux_vh}) and Eq.\,(\ref{eq:mechflux_vlambda}), respectively. For comparison, \citet{2019NatCo..10.3504L} derive, from observations alone, a lower limit of the average mechanical energy flux from swirls of $33$ to $131\,{\rm W}\,{\rm m}^{-2}$ at a height of approximately $z=1000\,{\rm km}$.
        
    %
    %%%%%%%%%%%%%%%%%%%%%%%%%%%%%%%%%%%%%%%%%%%%%%%%%%%%%%%%%%%%%%%%%%%%%%%%%%%%%%
    %    
    
    \subsubsection{Mean energy transport}\label{susubbsec:energetics2} 
    \begin{figure}
        \resizebox{\hsize}{!}
        {\includegraphics{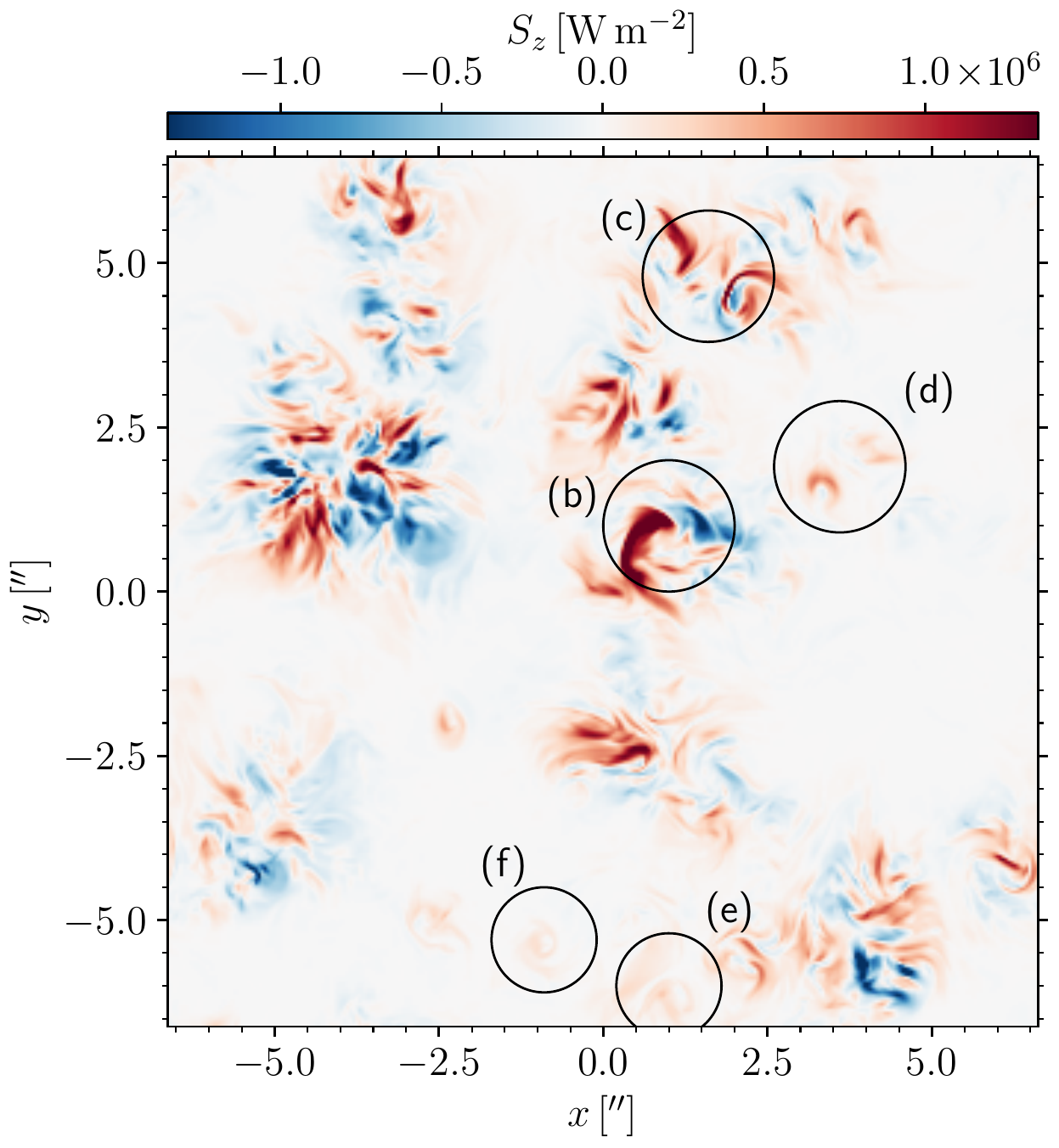}}
        \caption{
        Vertical component of the Poynting flux vector $S_z$ over the entire computational domain at $z=700\,\mathrm{km}$ and $t = 5800\,{\rm s}$. Black circles indicate the location of some of the swirl events listed in Table\,\ref{tab:ListSupplementarySwirls} also shown in Fig.\,\ref{fig:Movie_IcoBin5}.
        }
        \label{fig:z700_t5860}
    \end{figure}
    \begin{figure}
        \resizebox{\hsize}{!}
        {\includegraphics{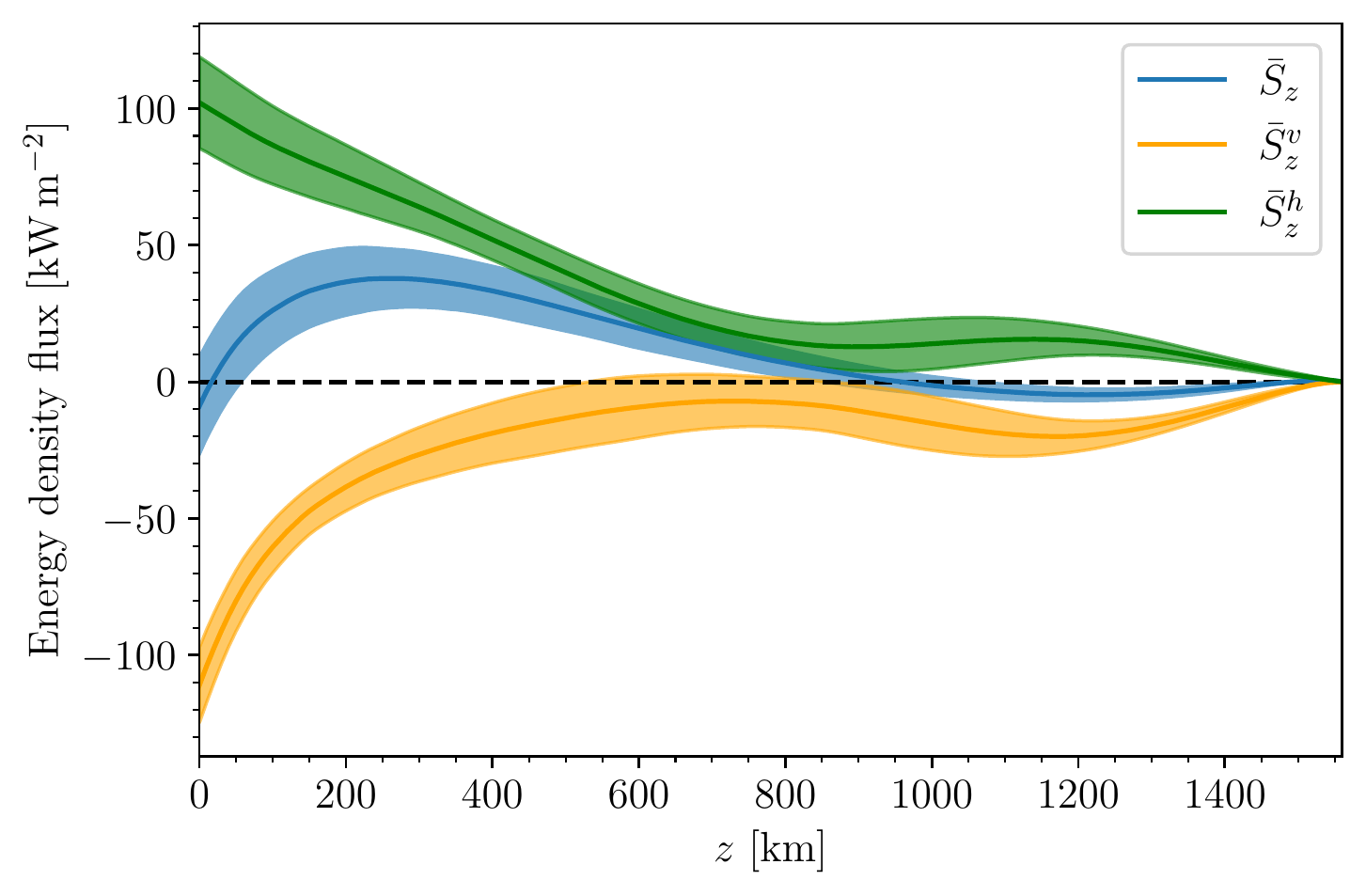}}
        \caption{
        Mean vertical component of the Poynting flux vector $S_z$ (blue) and of the terms $S_z^{\rm h}$ (green) and $S_z^{\rm v}$ (orange) as a function of height $z$. The spatial mean is obtained by averaging over the horizontal cross sections of the full computational domain of $9.6\times 9.6\,{\rm Mm}^2$, while the temporal mean is computed from $21$ different time instants spread over the time period of the high-cadence series of $15\,{\rm min}$. The shaded areas correspond to the standard deviation obtained from the temporal variation.
        }
        \label{fig:Av_PoyntingFlux}
    \end{figure}
    Looking at the mean net Poynting flux across the full computational domain of $9.6\times 9.6\,{\rm Mm}^2$, it turns out that this flux is much larger than the extrapolation from the single, relatively weak, isolated swirl event of Fig.\,\ref{fig:SingleSwirlEvent}, and that the bulk of energy transport is delivered by the more complex events of superposition of swirls (see Appendix\,\ref{app:superposition_of_swirls}).
    This situation is illustrated with Fig.\,\ref{fig:z700_t5860}, which shows the vertical component of the Poynting flux on the horizontal cross section at $z=700\,{\rm km}$ for the same time instant as Fig.\,\ref{fig:Movie_IcoBin5}, that is,
    $t = 5800\,{\rm s}$. In this figure, the largest single contribution to the mean vertical Poynting flux of $18.4 \, {\rm kW}\,{\rm m}^{-2}$ comes from event (b) near the center of the field of view with a peak Poynting flux of $2123.7\,{\rm kW}\,{\rm m}^{-2}$. Comparing it with  Fig.\,\ref{fig:Movie_IcoBin5}, it can be seen that this event starts from a relatively large magnetic footpoint patch with a complex but clearly swirling motion in the bin-5 intensity (see the animation of Fig.\,\ref{fig:Movie_IcoBin5}). For comparison, event (f) corresponds to the single, isolated swirl of Fig.\,\ref{fig:SingleSwirlEvent}.
    
    Figure \,\ref{fig:Av_PoyntingFlux} shows the temporal and spatial mean of the vertical component of the Poynting flux $S_{z}$ together with the terms $S_z^{\rm v}$ and $S_z^{\rm h}$ as a function of height $z$. The shaded area surrounding each curve indicates the standard deviation of the temporal variation over the time period of $15\,{\rm min}$ of the high-cadence simulation\footnote{We have also created an equivalent plot for the duration of $2\,{\rm h}$ for which we had snapshots available with a cadence of $4\,{\rm min}$. The result is very similar to Fig.\,\ref{fig:Av_PoyntingFlux}.}. First we observe that all curves converge to zero flux at the upper boundary, which is a consequence of the top boundary condition for the magnetic field. Therefore, magnetic disturbances are reflected at the top boundary with the consequence that the net vertical Poynting flux vanishes in the upper part of the atmosphere\footnote{This limitation could be lifted in a future simulation run with a more relaxed top boundary condition for the magnetic field; for example using the condition ${\rm d}\boldsymbol{B}/{\rm d}z = 0$, or matching to a potential field configuration.
    Varying boundary conditions would possibly lead to deviations in the energy fluxes surpassing the standard deviation shown in Fig.~\ref{fig:Av_PoyntingFlux}.}.
    Second, the contribution due to horizontal motions (including swirls) is always positive, while vertical motions contribute negatively to the Poynting flux. Above $z \approx 1000\,{\rm km}$, the negative contribution is even slightly dominating, possibly due to the boundary condition of vertical magnetic field at the top, which favors vertical motions in the upper part of the atmosphere. In this height range, $\beta < 0.1$ in regions of significant Poynting flux so that the top boundary condition strongly influences the magnetic field down to  $z \approx 1000\,{\rm km}$ because of the stiffness of this field. However, at the base of the chromosphere, which we take here again to be located at $z=700\,{\rm km}$, the mean flux is $12.8\pm6.5\,{\rm kW}\,{\rm m}^{-2}$, which, for a comparison, would be
    sufficient to compensate the radiative losses of $4.3\,{\rm kW}\,{\rm m}^{-2}$ in the chromosphere. 
    
    We also notice that at this height almost all the vertical Poynting flux is carried by the horizontal motions of the magnetized plasma ($S_z \approx S_z^{\rm h}$), that is, largely by swirling motions. This result agrees with the findings reported by \citet{2020ApJ...894L..17Y}. % Yadav et al, 2020
    Moreover, they demonstrate that the largest contribution to the Poynting flux comes from small-scale vortices that cluster to form the larger, observable swirls. This is consistent with our finding that superposition of swirls can carry large amounts of magnetic energy as in the case of event (b) of Fig.\,\ref{fig:z700_t5860}.
    This large magnetic energy flux is tantalizing, considering the findings of 
    \citet{2005Natur.435..919F} %Fossum & Carlsson (2021) 
    that high-frequency acoustic waves are not sufficient to heat the solar chromosphere and that the emission from the middle and upper chromosphere must be balanced by processes related to the magnetic field: a problem that is still unsolved 
    \citep{2019ARA&A..57..189C}. % Carlsson et al. (2019)
    \citet{2012A&A...541A..68M}  % Moll et al (2012)
    showed with the help of simulations similar to the present ones that Ohmic dissipation of magnetized swirling motions induce substantial local heating, although mostly in the photosphere. 
    
    However, a word of caution is in order here. The initial condition of a homogeneous vertical magnetic field together with the periodic lateral boundary conditions, of which the simulation keeps a memory for all times, favor the formation of vertically extended vortices. These simulations may adequately represent regions with a predominantly vertically directed magnetic field such as magnetic network patches or plage regions. With a more heterogeneous magnetic field, such as that generated by a turbulent dynamo in very quiet regions, magnetically induced swirls are probably less numerous and hence the Poynting flux less vigorous.

% 
%%%%%%%%%%%%%%%%%%%%%%%%%%%%%%%%%%%%%%%%%%%%%%%%%%%%%%%%%%%%%%%%%%%%%%%%%%%%%%%%%%
%%%%%%%%%%%%%%%%%%%%%%%%%%%%%%%%%%%%%%%%%%%%%%%%%%%%%%%%%%%%%%%%%%%%%%%%%%%%%%%%%%
%

\section{Summary and conclusions}\label{sec:conclusions}

    We carried out numerical radiation MHD simulations of the solar atmosphere with the purpose of studying the origin and propagation of vortical motions that occur in conjunction with magnetic fields and bear great resemblance to observed chromospheric swirls. The simulations are of fairly high spatial and temporal resolution and reach from the convectively unstable subsurface layers to the chromosphere. For the analysis, we mostly use a run that started with a homogeneous vertical magnetic field of a strength of $50\,{\rm G}$, but use for comparison also runs that started with a magnetic field of $200\,{\rm G}$ and without magnetic field.

    Looking first at the distribution of swirling strength, which is a measure for the vigor of swirling motions, we find a drastic difference between the simulations with and those without a magnetic field. The simulations with magnetic field, which show the usual funnel-shaped magnetic flux concentrations in the photosphere, have the swirling motion strongly concentrated within these low $\beta$ funnels and aligned with the magnetic field. Different from that, the simulation without magnetic field possesses more arch-shaped vortex tubes in the photosphere and almost isotropically distributed vortices in the chromosphere. These results agree with and confirm earlier findings of 
    \citet{2011A&A...533A.126M} % Moll et al. (2011)
    and \citet{2012ASPC..456....3S}. % Steiner & Rezei (2012)

    Beyond this relation between magnetic field and swirling motion, we find a tight relationship between swirling motions and perturbations in the magnetic field of intense magnetic flux tubes. In particular, we show with the help of a statistical analysis that the swirling strength and the magnetic swirling strength are highly correlated in the solar atmosphere. While swirling strength indicates a vortex in the plasma, the magnetic swirling strength indicates a twist in the magnetic lines of force. The latter typically occurs in the presence of torsional Alfvén waves. Therefore, the tight correlation between these two quantities is a first indication of the relationship between swirling motions and torsional Alfvén waves in the simulated solar atmosphere.

    We then analyze one particular swirl event that was identified in the numerical simulations (see Fig.\,\ref{fig:SingleSwirlEvent}). The well developed chromospheric swirl is colocated with a positive polarity magnetic BP in an intergranular lane of the deep photosphere and presents a twist in the magnetic field. Therefore, it can be identified as a magnetic swirl. Using time-distance diagrams to study its evolution in time and altitude, we find three pieces of evidence of the Alfv\'enic nature of this magnetic swirl. 
    First, the vortical motions and the magnetic perturbations, in the upwardly directed (positive polarity) magnetic field, have opposite orientation; second, the swirl propagates upward with the local Alfv\'en speed and, third, the driving force that is responsible for such propagation is the magnetic tension. These are all characteristics of torsional Alfv\'en waves. However, we do not observe an oscillatory wave, but instead pulses of unidirectional vortical motions and twisted magnetic field lines. From all that, we conclude that the swirl subsists in an upwardly propagating torsional Alfv\'en pulse. The other eight swirls listed in Table\,\ref{tab:ListSupplementarySwirls} also share these characteristics and are therefore also identified as torsional Alfvén pulses.
    A similar conclusion was also reached by 
    \citet{2019NatCo..10.3504L} % Liu et al. 2019
    from observations and modeling. The novelty of the present work is that these Alfv\'en pulses are not deliberately excited but naturally arise from a self-consistent numerical solution of the MHD equations  and radiative transfer.

    Regarding the origin of the Alfv\'en pulse, we cannot say with certainty if it is predominantly hydrodynamic or magnetic. From the analysis based on the swirling equation, we find that magnetic baroclinicity in the top convection zone and the deep photosphere is responsible for the creation of the swirl. Hence, magnetic dynamics alone would launch the pulse.
    However, the present time-distance analysis is based on a single, vertical slit through the center of the swirl, which does not follow the temporal evolution of the swirl nor its geometrical shape. This can lead to misinterpretations of the diagrams, especially in the chromosphere where the plasma flows are fast. %For example, a vortex that is simply advected out of the distance slit will look like being destroyed, while a swirl that is advected into the distance slit can be mistaken as the propagation of a swirl. 
    Therefore, it is necessary to consider the full three-dimensional structure of the individual terms of the swirling equation as a function of time in order to make progress. Such a three-dimensional analysis was beyond the scope of the present paper, but it can be expected to reveal whether the magnetic baroclinicity in the center of the swirl is caused by magnetic effects alone (e.g., magnetic diffusion or an MHD instability such as that responsible for quiet Sun Ellerman bombs) or by hydrodynamic effects in the surroundings of the magnetic flux concentration (e.g., by the bathtub effect or the granular buffeting).

    The upwardly propagating torsional Alfv\'en pulses carry a substantial amount of magnetic and mechanical energy from the photosphere to the outer layers of the atmosphere. At the top boundary of the computational domain, the Poynting flux vanishes because of the adopted boundary condition that forces the field to become strictly vertical. Therefore, we limited our analysis to the base of the chromosphere for which we took here the level $z=700$\,km. 

    For the swirl of Fig.\,\ref{fig:SingleSwirlEvent}, we evaluate at this height level a mean Poynting flux in the upward direction of $S_z = 34.5\,\mathrm{kW}\mathrm{m}^{-2}$. This flux is generated by the horizontal motions of the magnetized plasma alone and is therefore an effect of the horizontally swirling motion. Taking the observed frequency of chromospheric swirls from 
    \citet{2019NatCo..10.3504L} % Liu et al. (2019)
    and assuming they were all similar to the swirl of Fig.\,\ref{fig:SingleSwirlEvent}, we obtain a mean Poynting flux of $\bar{S}_z = 1.2\,{\rm kW}\,{\rm m}^{-2}$ and for the corresponding mean mechanical energy flux we find an upper limit of $\bar{F}_{z} = 218 - 544 \,{\rm W}\,{\rm m}^{-2}$. This energy flux could source the energy budget of the higher layers, but it is not enough to compensate alone for radiative losses at chromospheric level. 

    However, the swirl of Fig.\,\ref{fig:SingleSwirlEvent} is relatively weak; when evaluating the average Poynting flux across the full computational domain, still at a height of $z=700\,{\rm km}$, we obtain 
    $12.8\pm6.5\,{\rm kW}\,{\rm m}^{-2}$. This large amount mainly stems from more complex events of superposition of swirls, which originate in relatively large, complex, and intense magnetic footpoints.
    Assessing these numbers one should bear in mind that the initial condition of a homogeneous, vertical magnetic field favors the formation of vertically extended vortices and may be rather applicable to regions with a predominantly vertically directed magnetic field such as network and plage regions than to quiet Sun regions.

    Would it be possible to directly observe the Alfv\'enic nature of chromospheric swirls? We believe that it should be feasible with high resolution and high sensitivity spectropolarimetry in two spectral lines, preferably sampling the photosphere and the chromosphere. The challenge would consist in measuring the transverse magnetic field accurately enough to prove the twist of the magnetic field relative to the vortex motion that would be detected with the help of LCT. The two height levels would inform about the propagation speed of the swirl. High photon flux is needed for accurately measuring the transverse Zeeman effect, which requires a large aperture solar telescope such as the D.\,K.\,Innouye Solar Telescope (DKIST).
%
%%%%%%%%%%%%%%%%%%%%%%%%%%%%%%%%%%%%%%%%%%%%%%%%%%%%%%%%%%%%%%%%%%%%%%%%%%%%%%%%%%
%%%%%%%%%%%%%%%%%%%%%%%%%%%%%%%%%%%%%%%%%%%%%%%%%%%%%%%%%%%%%%%%%%%%%%%%%%%%%%%%%%
%

\begin{acknowledgements}
      AFB wishes to acknowledge support by the Swiss National Science Foundation (SNF) through grant 200021\_189180 (Solar Orbiter STIX), JRCC support by SNF under grant ID 200020\_182094, and FC support through the CHROMATIC project (2016.0019) of the Knut and Alice Wallenberg foundation. Special thanks are addressed to L.\,Harra who directed AFB's Master's thesis from which the present work derives. This work has profited from discussions with the team of K.\,Tziotiou and E.\,Scullion (conveners) ``The Nature and Physics of Vortex Flows in Solar Plasma'' at the International Space Science Institute (ISSI) and with the Waves in the Lower Solar Atmosphere (WaLSA; www.WaLSA.team) team (S.\,Jafarzadeh, convener), which is supported by the Research Council of Norway (project number 262622).
      Simulations were carried out at the Swiss National Supercomputing Centre (CSCS) under project ID s560 with support from SNF under grant ID 200020\_157103. Thanks are also extended for valuable input by G.\,Vigeesh and P.\,Rajaguru, and for very helpful and encouraging comments by an anonymous referee. 
      
\end{acknowledgements}

%--------------------------------------------------------------------


% for the bibliography, at the end
\bibliographystyle{aa} % style aa.bst
\bibliography{aanda.bib} % your references Yourfile.bib

%
%%%%%%%%%%%%%%%%%%%%%%%%%%%%%%%%%%%%%%%%%%%%%%%%%%%%%%%%%%%%%%%%%%%%%%%%%%%%%%%%%%
%%%%%%%%%%%%%%%%%%%%%%%%%%%%%%%%%%%%%%%%%%%%%%%%%%%%%%%%%%%%%%%%%%%%%%%%%%%%%%%%%%
%

\begin{appendix}
    
\section{Strength and distribution of vortices}\label{app:comparison}

    This appendix compares three different CO$5$BOLD simulations: one purely hydrodynamic without magnetic field and two MHD ones with an initial homogeneous, vertical magnetic field of $50\,{\rm G}$, and $200\,{\rm G}$. We shall refer to them as Hydro, MHD $50\,{\rm G}$, and MHD $200\,{\rm G}$ hereafter. The MHD $200\,{\rm G}$ run corresponds to \texttt{d3gt57g44v200fc} of
    \citet{doi:10.13097/archive-ouverte/unige:115257}, % Calvo (2018)
    while the Hydro run to \texttt{d3gt57g44roefc} of \citet{2016A&A...596A..43C} and %Calvo et al. (2016)
    \citet{doi:10.13097/archive-ouverte/unige:115257}. % Calvo (2018)
    
\begin{figure*}%[!]
        \includegraphics[width=\hsize]{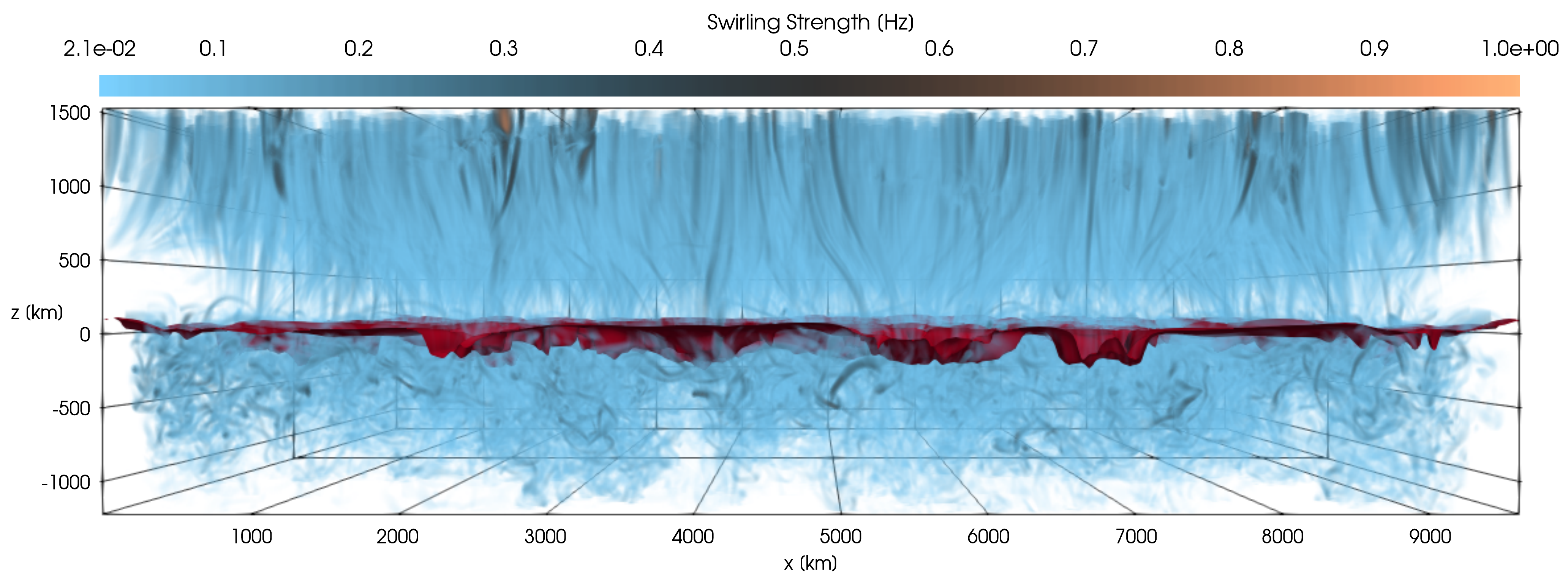}
        \label{fig:SwirlsDistribution_v200}
        \\
        \includegraphics[width=\hsize]{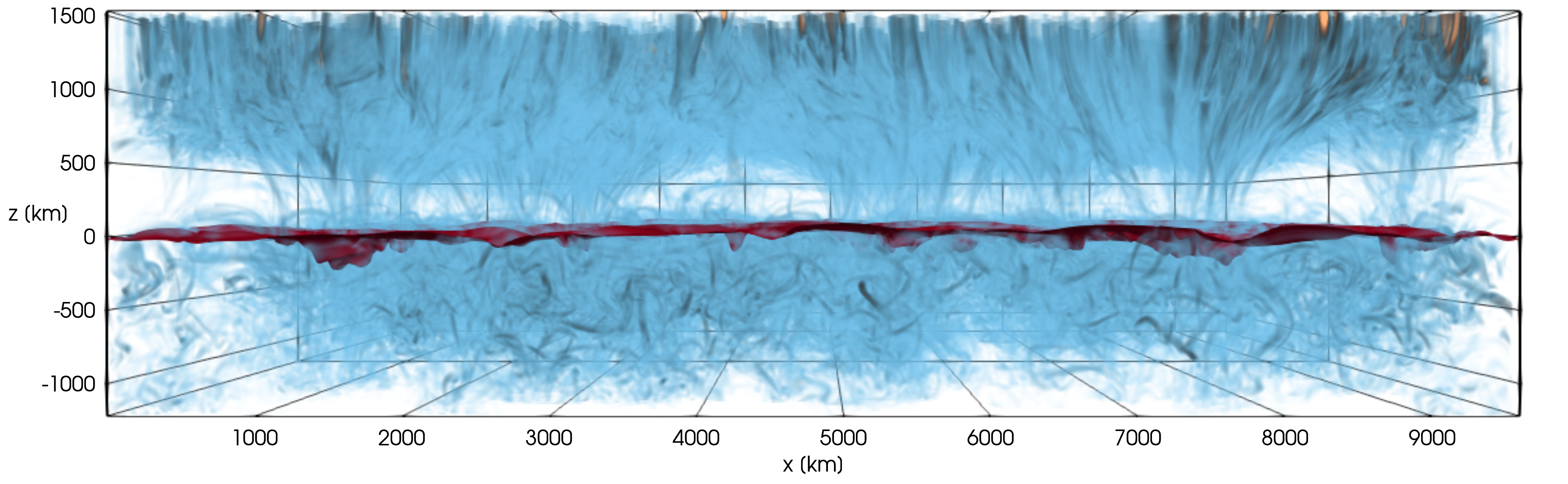}
        \label{fig:SwirlsDistribution_v50}
        \\
        \includegraphics[width=\hsize]{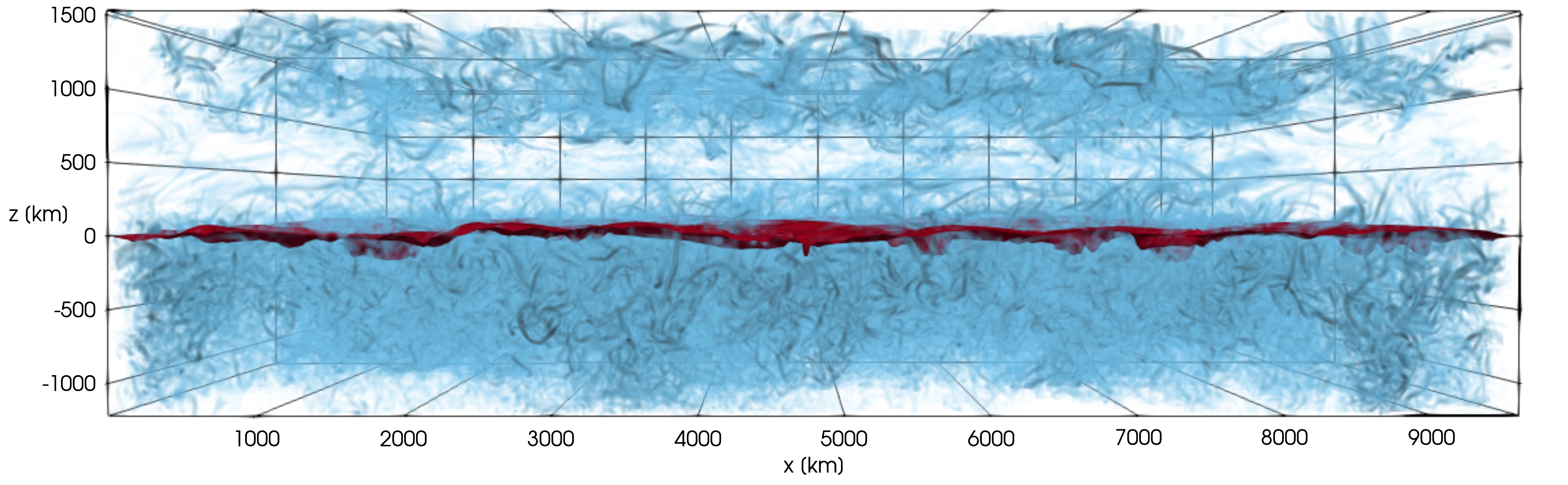}
        \label{fig:SwirlsDistribution_hydro}
        \caption{Volume rendering of the swirling strength seen from the side for three different simulations at arbitrary time instants: (\emph{top}) MHD simulation with an initial vertical, homogeneous magnetic field of $200\,{\rm G}$; (\emph{middle}) MHD simulation with an initial field of $50\,{\rm G}$; 
        (\emph{bottom}) purely hydrodynamical simulation. The red surface indicates the $\tau_{500}=1$ surface. Swirls with a period larger than $10\,{\rm min}$ ($\lambda < 2.09 \times 10^{-2}\,{\rm Hz}$) have been neglected and values above $1\,{\rm Hz}$ are saturated.
        }
        \label{fig:SwirlsDistribution_ThreeDifferentSimulations}
    \end{figure*}

    Figure\,\ref{fig:SwirlsDistribution_ThreeDifferentSimulations} shows instants of the swirling strength $\lambda$ for each of the three simulations: from top to bottom, MHD $200\,{\rm G}$,  MHD $50\,{\rm G}$, and Hydro. The $\tau_{500}=1$ surface is displayed in red. Swirls with $\lambda < 2.09 \times 10^{-2}\,{\rm Hz}$, that is, with a period larger than $10\,{\rm min}$, are neglected, while strong swirls with $\lambda > 1.0\,{\rm Hz}$ are saturated.
    Comparing the three panels of Fig.\,\ref{fig:SwirlsDistribution_ThreeDifferentSimulations}, one recognizes that below the surface of $\tau_{500}=1$, the three simulations share a nearly isotropic distribution of swirls. This behavior was also found by 
    \citet{2011A&A...533A.126M}. %Moll et al 2011 
    These swirls are obviously caused by the turbulent plasma motions in the surface layers of the convection zone. 
    However, the Hydro simulation shows a higher density of swirls than the MHD simulations. This is because the magnetic fields hamper the formation of vortices in the convective layers of the simulation.
    
    The photosphere of the MHD $200\,{\rm G}$ time instant harbors an almost homogeneous structure of vertically oriented swirls and only a few swirl arches are present near the $\tau_{500}=1$ surface.  In the MHD $50\,{\rm G}$ case, the vertical swirls are clustered in funnel-shaped structures and the arches are  more prominent than in MHD $200\,{\rm G}$. Different from these two cases, vertical swirls are essentially absent in the Hydro simulation, while swirl arches are omnipresent in the photosphere.
    Again, these results well agree with and confirm earlier results of
    \citet{2011A&A...533A.126M, 2012A&A...541A..68M}. % Moll et al 2011 & 2012  
    
    A similar behavior is observed in the chromosphere: the stronger the initial magnetic field, the more abundant the vertical swirls are. However, we observe that the strongest swirls appear at the top of the MHD $50\,{\rm G}$ simulation. This suggests that too strong magnetic fields can possibly inhibit the formation of strong vertical vortices in the chromosphere. In the Hydro simulation, we see a quasi-isotropic distribution of swirls at chromospheric levels, which are most probably related to hydrodynamic shocks taking place in this region of the field-free solar atmosphere \citep[see, e.g.,][]{2012A&A...541A..68M}. %Moll et al 2012
    Therefore, not all swirls found at chromospheric levels are necessarily related to the magnetic field, even though the magnetism clearly plays a cardinal role in the generation of vertical vortices.

    \begin{figure*}
        \resizebox{\hsize}{!}
        {\includegraphics{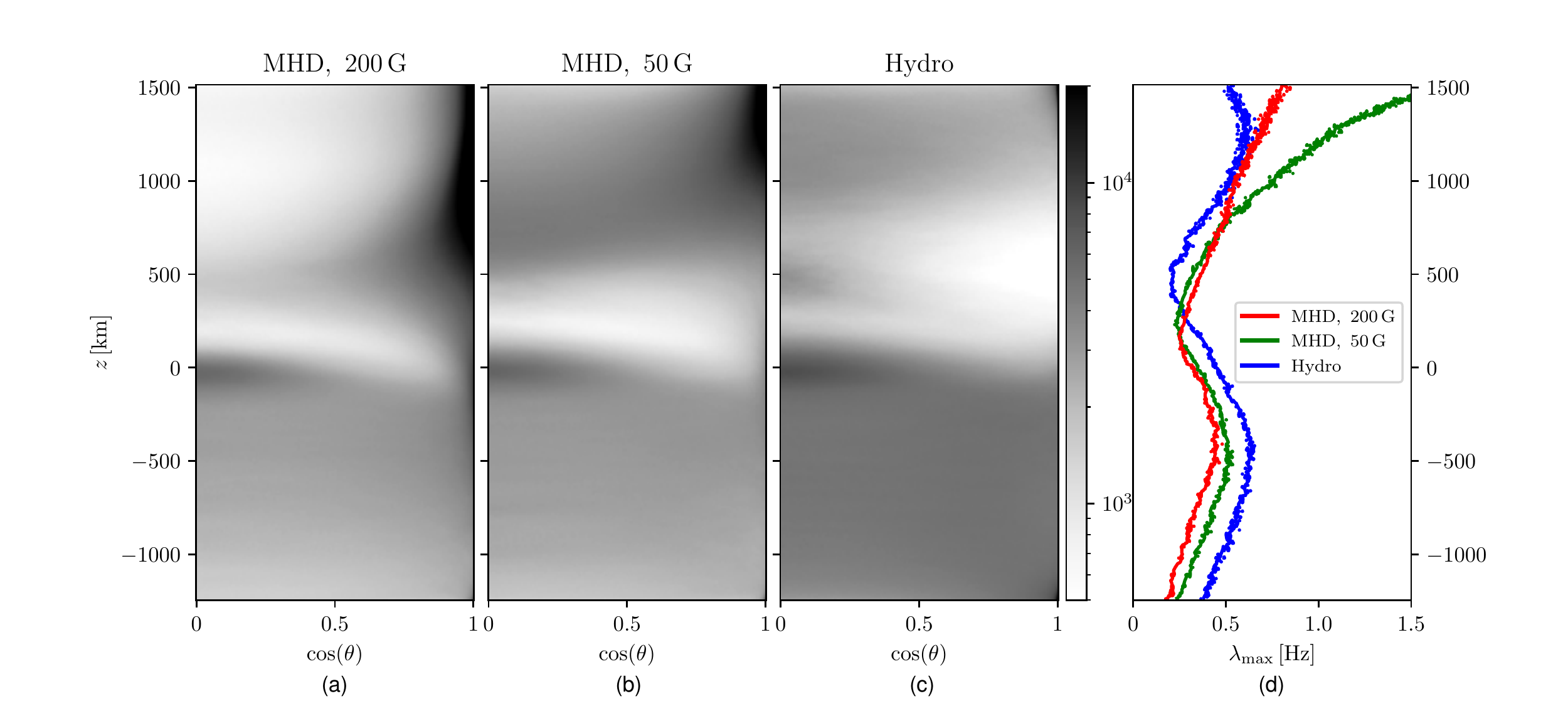}}
        \caption{Distribution of the inclination $\cos(\theta)$ of the swirling strength vector as a function of altitude $z$ for three different simulations: (\emph{a}) MHD simulation with an initial vertical, homogeneous magnetic field of $200\,{\rm G}$, (\emph{b}) MHD simulation with an initial field of $50\,{\rm G}$, and (\emph{c}) purely hydrodynamical simulation. The inclination $\theta$ refers to the angle between the swirl axis and the vertical direction. Horizontally oriented swirls are binned close to $\cos(\theta)=0$, while vertically oriented swirls are close to $\cos(\theta)=1$. The maximum value of swirling strength $\lambda_{\rm max}$ at each height $z$, for each simulation, and with a minimum of $10$ occurrences per bin, is shown in (\emph{d}).
        The occurrences are means over $5$ different time instants of $960\times960\times280$ data points for each simulation. The bin sizes are $\Delta z = 10\,{\rm km}$ and $\Delta \cos(\theta) = 3.62\times10^{-3}$. Swirls with $\lambda < 2.09 \times 10^{-2}\,{\rm Hz}$ are neglected.}  
        \label{fig:SwirlsDistribution_ThreeSimOrientations}
    \end{figure*}

    The above more qualitative assessment is confirmed by the quantitative analysis of Fig.\,\ref{fig:SwirlsDistribution_ThreeSimOrientations}, which shows for each simulation a bi-dimensional histogram of the inclination angle $\cos(\theta)$ of the swirling strength vector at each height $z$. The angle $\theta$ is defined through
    \begin{equation}
        \cos(\theta) = \frac{\boldsymbol{\lambda}\cdot\hat{\boldsymbol{e}}_z}{|\boldsymbol{\lambda}|} = \frac{\lambda_z}{\lambda}\,, \nonumber
    \end{equation}
    being the angle between the swirling strength vector $\boldsymbol{\lambda}$ and the unit vector in the vertical direction $\hat{\boldsymbol{e}}_z$.
    Therefore, horizontally oriented swirls are binned close to $\cos(\theta)=0$, while vertically oriented swirls are close to $\cos(\theta)=1$.
    We choose to plot the histogram of $\cos(\theta)$ instead of $\theta$ in order to easily recognize isotropy. In this way, an isotropic angle distribution at a certain height $z$ is characterized by a flat distribution in the histograms of Fig.\,\ref{fig:SwirlsDistribution_ThreeSimOrientations}.   
    
    In the convection zone, that is, for $z\lesssim 0\,{\rm km}$, all three simulations show a flat distribution, which confirms the nearly isotropic distribution of swirls in that region. Also, as estimated before, there are on average more swirls in the Hydro simulation (c) than in the MHD $50\,{\rm G}$ (b) and the MHD $200\,{\rm G}$ (a) simulations. Near the optical surface, that is, around $z=0\,{\rm km}$, there is a sudden break of isotropy. Two families of vortices can be spotted. First, a cluster of horizontally oriented swirls ($\cos(\theta) \approx\! 0$), visible in all three panels: They stem from the arches of swirling strength that are seen to populate the near surface layers. Second, a cluster of vertically oriented swirls ($\cos(\theta) \approx\! 1$) present only in the MHD simulations. Both groups of swirls were already seen in Fig.\,\ref{fig:SwirlsDistribution_ThreeDifferentSimulations}. We notice that the number of horizontally oriented swirls slightly increases as the initial magnetic field decreases. The opposite happens for the vertically oriented swirls, which are more numerous in the MHD $200\,{\rm G}$ simulation than in MHD $50\,{\rm G}$.
    Higher up, the vortices are essentially uniquely vertical in the case of MHD $200\,{\rm G}$ (a), still predominantly vertical in MHD $50\,{\rm G}$ (b) and almost isotropically distributed in the Hydro simulation (c). This confirms once again the visual perception given by Fig.\,\ref{fig:SwirlsDistribution_ThreeDifferentSimulations}.
    
    Caution is indicated concerning the regions close to the top and to the bottom boundaries of the computational domain. For the MHD simulations, the magnetic field is forced to be vertical at the top and bottom boundaries. In the purely hydrodynamic case, this condition is not present. Nevertheless, a large number of weak, vertical swirls are encountered at the top and bottom boundaries of all three simulations. Most of them have been removed by the application of the threshold, $\lambda = 2.09 \times 10^{-2}\,{\rm Hz}$, in Figs.\,\ref{fig:SwirlsDistribution_ThreeDifferentSimulations} and \ref{fig:SwirlsDistribution_ThreeSimOrientations}. However, an over-density of vertical swirls is still present at the bottom boundary of panels (a), (b), (c), and at the top boundary of the Hydro simulation.
    
    Panel (d) of Fig.\,\ref{fig:SwirlsDistribution_ThreeSimOrientations} shows the maximum value of the swirling strength, $\lambda_{\rm max}$, as a function of $z$ for the three simulations. From it, we notice that in the convection zone the stronger the initial magnetic field, the weaker the swirls. However, the overall behavior is similar and the local peak is reached in all three cases around $z \approx\!- 500\,{\rm km}$. Thus, relatively strong vortices can develop in the convection zone, independently of the initial magnetic configuration of the simulation.
    
    In the photosphere, a local minimum in swirling strength is reached. It is a result of the slowdown of vortical motions in the rapidly expanding convective overshoots as a consequence of angular momentum conservation 
    \citep{1997A&A...328..229N}. % Nordlund et al. (1997)
    For the two MHD simulations, the minimum is reached in the lower half of the photosphere, while the Hydro simulation reaches its minimum at $z \approx\!500\,{\rm km}$. This difference can be attributed to strong arches of swirling strength  in the hydrodynamical simulation, which permeate the photosphere and extend precisely up to $z \approx\!500\,{\rm km}$, as it can be seen in Fig.\,\ref{fig:SwirlsDistribution_ThreeDifferentSimulations}. These arches are less developed in the MHD simulations, probably as a consequence of the magnetic canopy.
    
    Finally, looking at the photosphere and chromosphere, we observe that $\lambda_{\rm max}$ increases exponentially for the MHD $50\,{\rm G}$ simulation, while this growth is only linear for the MHD $200\,{\rm G}$ simulation. In the Hydro case instead, there is a growth only in the low chromosphere up to approximately $z\approx\!1200\,{\rm km}$. Above this height, the swirls become weaker again. Clearly, magnetic fields favor the generation of swirls. However, it seems that too strong magnetic fields can also suppress the formation of strong vortices. We surmise that there exists an optimal value of the initial magnetic field for which the strength of the vortices is maximal.
    
%
%%%%%%%%%%%%%%%%%%%%%%%%%%%%%%%%%%%%%%%%%%%%%%%%%%%%%%%%%%%%%%%%%%%%%%%%%%%%%%%%%%%%%
%%%%%%%%%%%%%%%%%%%%%%%%%%%%%%%%%%%%%%%%%%%%%%%%%%%%%%%%%%%%%%%%%%%%%%%%%%%%%%%%%%%%%
%

\section{Superposition of swirls}\label{app:superposition_of_swirls}
    Section\,\ref{subsec:alfvén_pulses} considers the torsional Alfvén pulses of a single swirl event. An important characteristic of that event is that its magnetic footpoint is small compared with others that can be found in the simulation and that it is quite isolated from the rest of magnetic flux concentrations. In this sense, the swirl of Fig.\,\ref{fig:SingleSwirlEvent} is an ideal, clear-cut case, which we have intentionally chosen for ease of analysis. Whenever the magnetic footpoint is larger or when it consists of several footpoints close together, as may be representative of a magnetic network patch, a filigree, or a magnetic knot, the corresponding intensity pattern in the chromosphere becomes more involved. These sorts of complex magnetic structures can be energetically more important since multiple swirls may coexist in there. In the light of this phenomenon, we introduce the concept of ``superposition of swirls.''
    
\begin{sidewaysfigure*}[p]
        \centering
        \includegraphics[width=23cm]{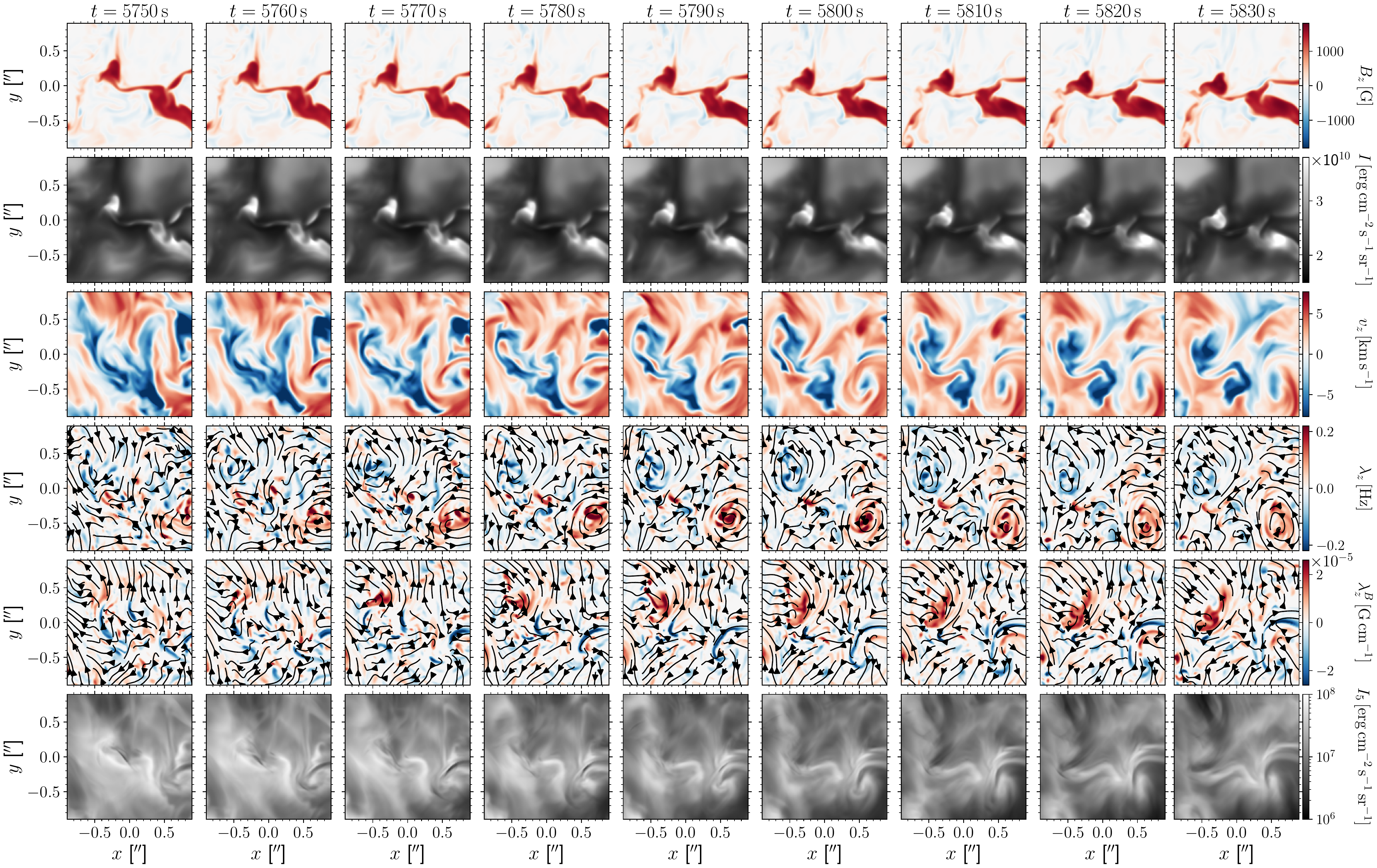}
        \caption{Time sequence of a superposition event from $t=5750\,\mathrm{s}$ to $t=5830\,\mathrm{s}$. \emph{From top to bottom row}: Vertical component of the magnetic field $B_z$ at $z=0\,\mathrm{km}$, continuum intensity $I$, vertical velocity $v_z$ at $z=700\,\mathrm{km}$, vertical component of the swirling vector $\lambda_z$ at $z=700\,\mathrm{km}$, vertical component of the magnetic swirling vector $\lambda^{B}_z$ at $z=700\,\mathrm{km}$, and the bin-5 intensity $I_5$. Maps of $\lambda_z$ and $\lambda_z^{B}$ also show the streamlines of the velocity field and the magnetic field projected into the horizontal plane at $z=700\,\mathrm{km}$, respectively.
        }
        \label{fig:SuperpositionSwirls}
    \end{sidewaysfigure*}

     Figures\,\ref{fig:SuperpositionSwirls} and \ref{fig:SuperpositionEvent_Appendix} both show the time sequence of a superposition of swirls in which the magnetic footpoint is distinctly larger and has a more complex structure than that of Fig.\,\ref{fig:SingleSwirlEvent}. The figures are organized in the same way as Fig.\,\ref{fig:SingleSwirlEvent}.
     Especially in the case of Fig.\,\ref{fig:SuperpositionSwirls}, the bin-5 intensity, $I_5$, does no longer show a clear swirling motion, as was the case in Fig.\,\ref{fig:SingleSwirlEvent}. However, this does not imply that no swirling motions are present. Indeed, the vertical component of the swirling vector clearly shows two distinct swirls rotating in opposite direction: a clockwise vortex in the top left corner of the field of view and a counterclockwise one in the bottom right corner. 
     
\begin{figure*}
        \resizebox{\hsize}{!}
        {\includegraphics{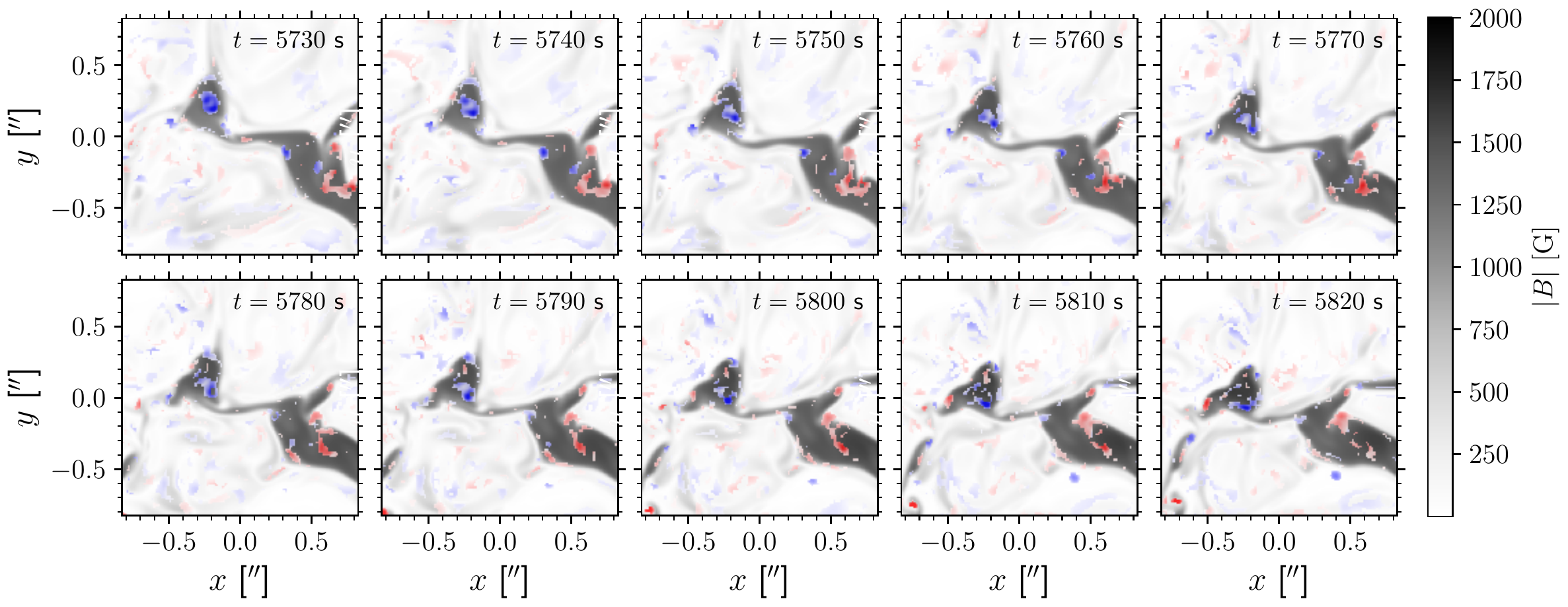}}
        \caption{Time sequence of the absolute magnetic field strength at $z=0$ (gray scale) in the same field of view as shown in Fig.\,\ref{fig:SuperpositionSwirls}. Plotted on top of the field strength is the vertical component of the swirling vector $\lambda_z$ in regions where $\lambda > 10^{-3}\,{\rm Hz}$ and $\lambda^{B} > 10^{-6}\,{\rm G}\,{\rm cm}^{-1}$. Red indicates counterclockwise and blue clockwise rotation, that is, positive and negative values of $\lambda_z$, respectively. }
        \label{fig:SuperpositionEvent_MagneticFieldEvolution}
    \end{figure*}

     In principle, the same analysis done in Sect.\,\ref{sec:results} also applies in the case of these two swirls. The more complex pattern of the bin-5 intensity can be explained by the geometry and dynamics of the large footpoint. To see this, Fig.\,\ref{fig:SuperpositionEvent_MagneticFieldEvolution} depicts the evolution of the magnetic field strength at $z=0\,{\rm km}$ for the same field of view as it is shown in Fig.\,\ref{fig:SuperpositionSwirls}. Each map is overlaid with the vertical component of the swirling vector in regions where $\lambda > 10^{-3}\,{\rm Hz}$ and $\lambda^{B} > 10^{-6}\,{\rm G}\,{\rm cm}^{-1}$. Here, red indicates positive (counterclockwise) swirling strength and blue stands for negative (clockwise) swirling strength. The distribution of the swirling strength in this region is fragmented since within the same magnetic structure several spots of swirling strength with different orientations occur. The reason for this is that the whole magnetic footpoint is not rigidly rotating, but possesses substructure that can have different relative motions and rotations. Consequently, different rotation directions within the same magnetic footpoint can be expected to occur. As the magnetic fields expand with height and start to fill the entire available volume at chromospheric levels, these individual rotational motions result in a superposition of different chromospheric swirls. (Provided that the vertical propagation effectively takes place, as is the case here.) However, a swirling motion detected at $z=0\,\mathrm{km}$ does not necessarily propagate up to the chromosphere. Some of these swirls may be the footpoints of vortex arches, they may be too weak to reach the chromosphere, or may become dominated by neighboring upwardly propagating perturbations.
    
    Figure\,\ref{fig:SuperpositionEvent_MagneticFieldEvolution} shows different patches of clockwise (negative) swirling strength within the left part of the magnetic footpoint. This then becomes manifest as a clockwise vortical motion in the plasma visible in the top left corner of $\lambda_z$ at $z=700\,{\rm km}$ from $t=5780\,{\rm s}$ onward (fourth row of Fig.\,\ref{fig:SuperpositionSwirls}). We also notice that the horizontal component of the positive polarity magnetic field has opposite direction with respect to the velocity field, indicating the torsional Alf\'enic nature of the pulse. Within the right part of the magnetic footpoint, we observe a dominance of patches of counterclockwise (positive) swirling strength. These become manifest as a counterclockwise vortical motion in the upper photosphere to lower chromosphere, which can be seen in the bottom right corner of $\lambda_z$ in Fig.\,\ref{fig:SuperpositionSwirls}. 
    
    The consequences for chromospheric observations can be guessed from the bin-5 intensity maps, which are distinctly different from the ones shown in Fig.\,\ref{fig:SingleSwirlEvent}. They still show traces of swirling motions, but they are less obvious than in the case of an unidirectional swirl event. The superposition of swirls is a ubiquitous phenomenon in our simulations since large footpoints are more frequent than small, isolated ones.
    The present interpretation is consistent with the findings of \citet{2020ApJ...894L..17Y}. Using MHD simulations, they studied vortex flows at different spatial scales and concluded that the observed large vortices likely consists of clusters of smaller swirls that are not resolved by the observations.

    In the case of superposed swirls, caution is indicated when analyzing time-distance diagrams. Because of the horizontal motion of the footpoints, neighboring swirls of opposite rotation may enter and leave the time-distance slit, giving the impression of an oscillatory torsional motion. Thus, when performing this type of analysis, care must be taken that the very same perturbation is traced along the distance slit in time. This effect could possibly explain the oscillatory, torsional motion detected by \citet{2013ApJ...776L...4S} % Shelyag et al., 2013
    from time-distance diagrams of the vorticity in the top region of their simulation domain.
    
    In conclusion, we interpret complex rotational motions above large footpoints as the superposition of multiple, unidirectional swirls, which are caused by separate, upwardly propagating, torsional Alfvén pulses. They, in turn, originate in the dynamics of substructure elements of the magnetic footpoint.
    
%
%%%%%%%%%%%%%%%%%%%%%%%%%%%%%%%%%%%%%%%%%%%%%%%%%%%%%%%%%%%%%%%%%%%%%%%%%%%%%%%%%%%%%
%%%%%%%%%%%%%%%%%%%%%%%%%%%%%%%%%%%%%%%%%%%%%%%%%%%%%%%%%%%%%%%%%%%%%%%%%%%%%%%%%%%%%
%

\section{Supplementary swirls}
\label{app:supplementary_swirls}
     \begin{table*}
        \caption{Swirl events manually detected in the simulation data.}
        \centering
        {
        \fontsize{10}{14}\selectfont
        \begin{tabular}{|lrrccll|}
        \hline
         \textbf{Label} & \multicolumn{2}{c}{\textbf{Coordinates}} & \textbf{Time interval} &  
         \textbf{Approximate size}& \textbf{Kind of swirl} & \textbf{Figure}\\
         &     $x\,\mathrm{[^{\prime\prime}]}$     &      $y\,\mathrm{[^{\prime\prime}]}$   & [s] &
         $\mathrm{[^{\prime\prime}]}$ &  &  \\ \hline
        (a) &    5.0     &    1.7      & 5520 - 5750 & 1.5 & Single swirl & - \\ \hline
        (b)                &    1.0     &    1.0      & 5700 - 5850 & 1.5   & Superposition event   & Fig.\,\ref{fig:SuperpositionEvent_Appendix} \\ \hline
        (c) &    1.6     &    4.8      & 5700 - 5900 & 1.5 & Superposition event & Fig.\,\ref{fig:SuperpositionSwirls}  \\ \hline
        (d) &    3.6     &    1.9      & 5760 - 6040 & 1.5 & Superposition event & - \\ \hline
        (e) &    1.0     &   -6.0      & 5770 - 5990 & 1.0 & Single swirl & Fig.\,\ref{fig:SingleSwirlEvent_Appendix} \\ \hline
        (f) &   -0.9     &   -5.3      & 5780 - 6050 & 1.4 & Single swirl & Fig.\,\ref{fig:SingleSwirlEvent}  \\ \hline
        (g)                &    3.7     &   -4.5      & 6000 - 6130 & 1.5   & Superposition event & - \\ \hline
        (h) &   -3.5     &    2.8      & 6270 - 6400 & 1.0 & Superposition event & - \\ \hline
        (i)                &    3.0     &    6.0      & 6280 - 6380 & 1.0   & Superposition event & - \\ \hline
        \end{tabular}
        }
        \tablefoot{
        All these events show upward propagation of close to plane parallel vortical motions, bearing the characteristics of torsional Alfvénic pulses.}
        \label{tab:ListSupplementarySwirls}
    \end{table*}
    \begin{figure*}
        \includegraphics[width=\hsize]{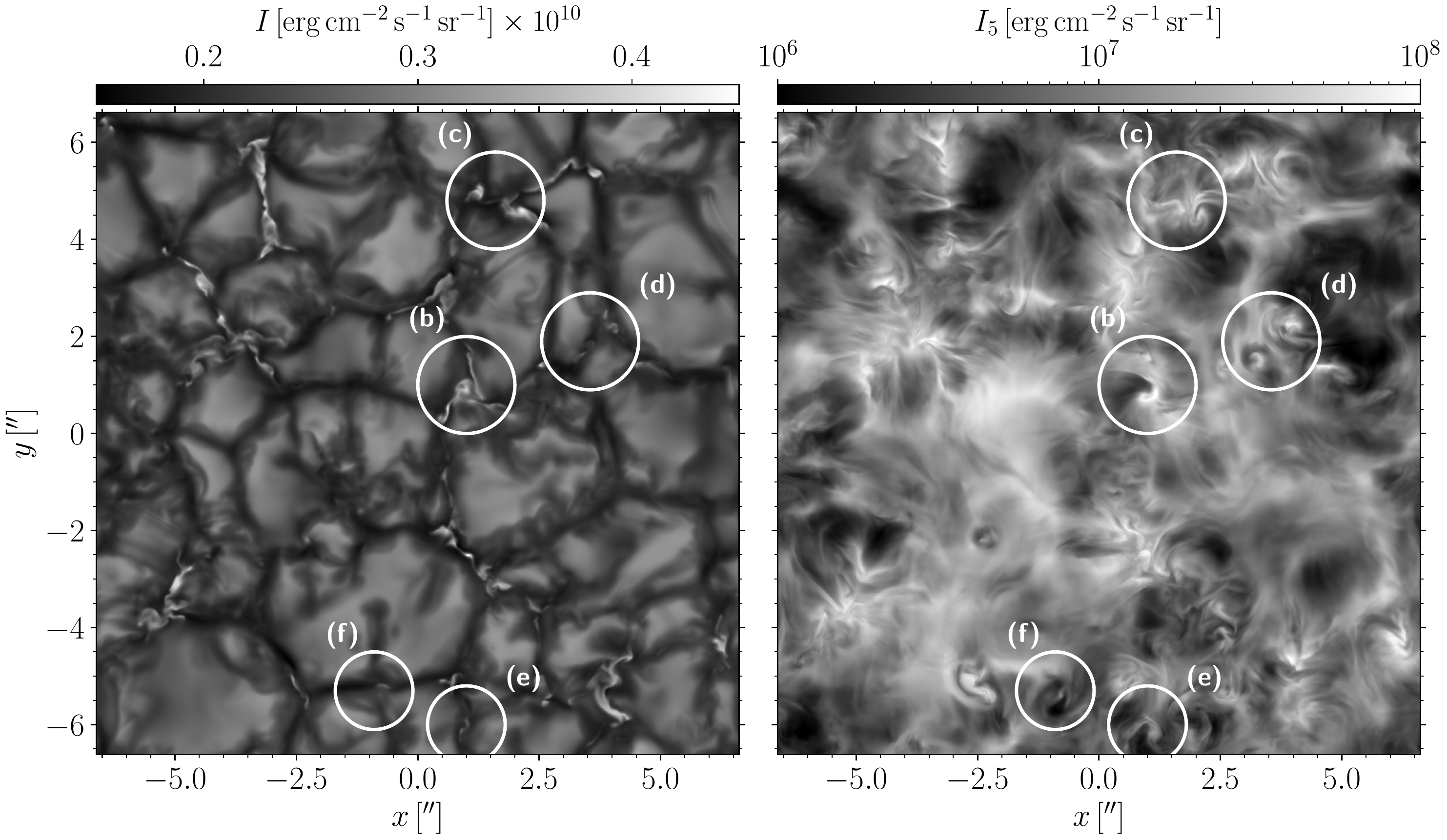}
        \caption{Time instant $t=5800\,{\rm s}$ showing (\emph{left}) the continuum intensity with the typical granulation pattern of the solar surface and (\emph{right}) the bin-5 intensity $I_5$ as a proxy for a chromospheric line core intensity. White circles indicate the location of some of the swirl events listed in Table\,\ref{tab:ListSupplementarySwirls}. An animation of this figure is available online.}
        \label{fig:Movie_IcoBin5}
    \end{figure*}
    \begin{sidewaysfigure*}[p]
        \centering
        \includegraphics[width=23cm]{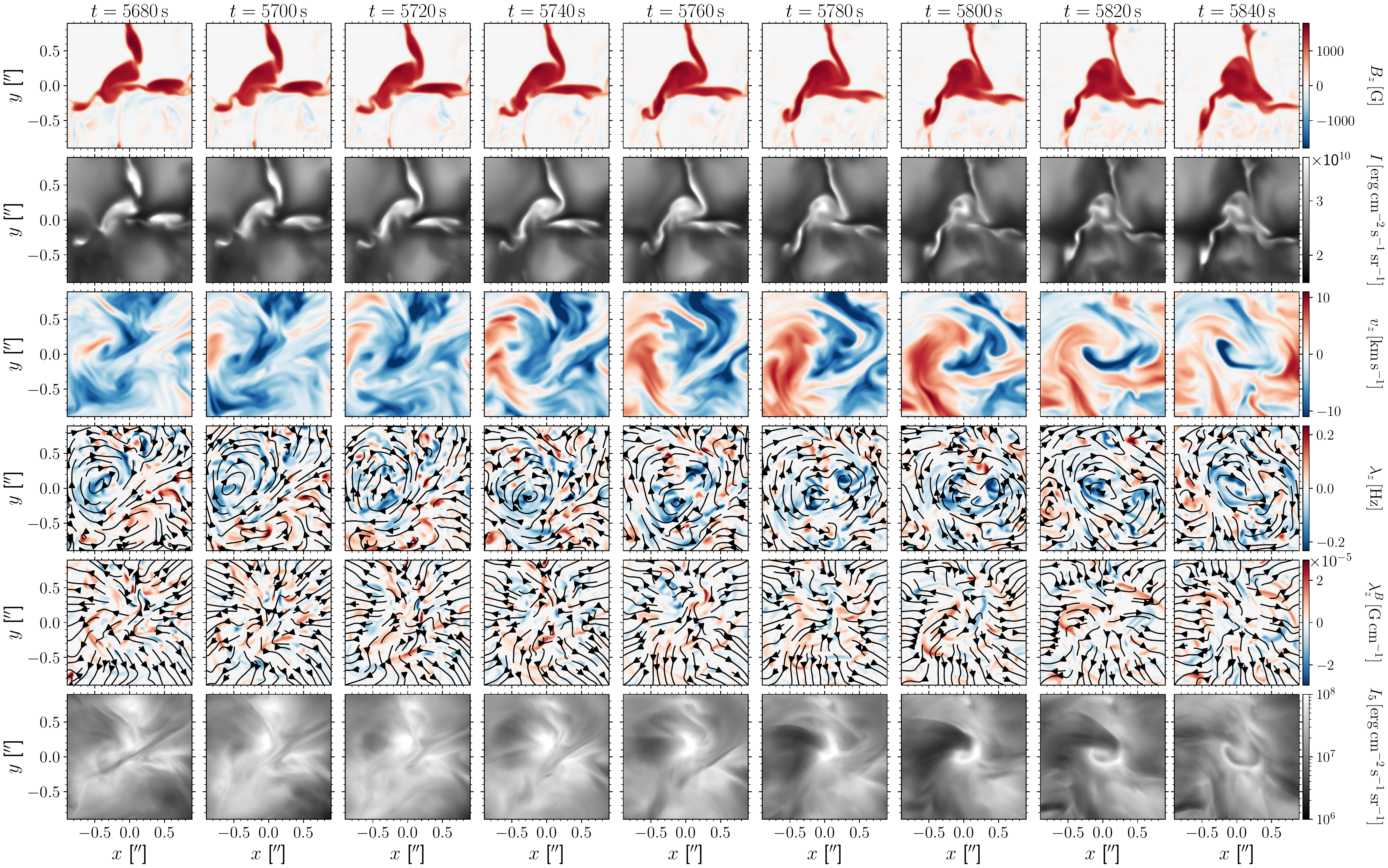}
        \caption{
        Time sequence of a superposition event from $t=5680\,\mathrm{s}$ to $t=5840\,\mathrm{s}$. \emph{From top to bottom row}: Vertical component of the magnetic field $B_z$ at $z=0\,\mathrm{km}$, continuum intensity $I$, vertical velocity $v_z$ at $z=700\,\mathrm{km}$, vertical component of the swirling vector $\lambda_z$ at $z=700\,\mathrm{km}$, vertical component of the magnetic swirling vector $\lambda^{B}_z$ at $z=700\,\mathrm{km}$, and the bin-5 intensity $I_5$. Maps of $\lambda_z$ and $\lambda_z^{B}$ also show the streamlines of the velocity field and the magnetic field projected into the horizontal plane at $z=700\,\mathrm{km}$, respectively.
        }
        \label{fig:SuperpositionEvent_Appendix}
    \end{sidewaysfigure*}
    \begin{sidewaysfigure*}[p]
        \centering
        \includegraphics[width=23cm]{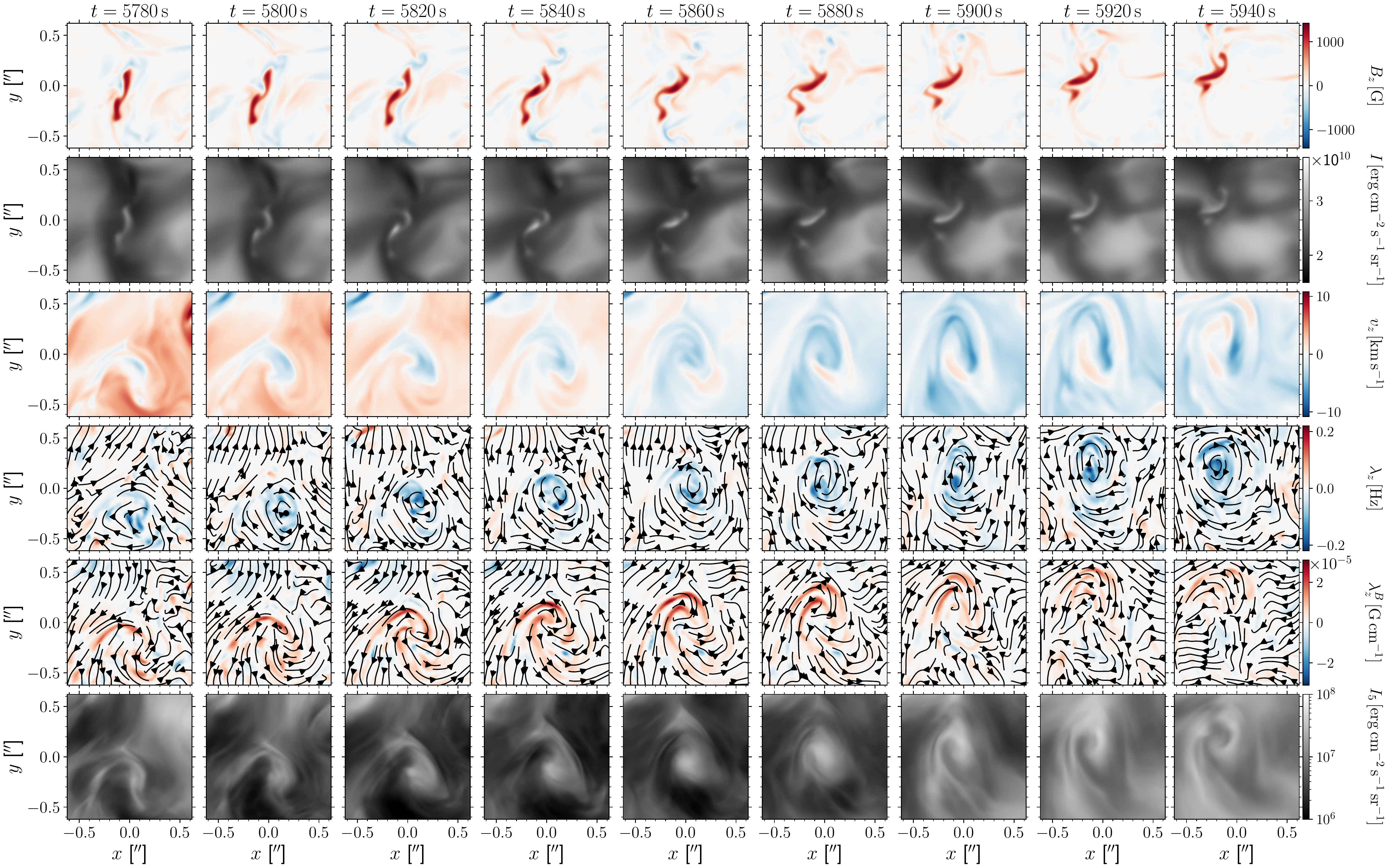}
        \caption{
        Time sequence of a single swirl event from $t=5780\,\mathrm{s}$ to $t=5940\,\mathrm{s}$. \emph{From the top to the bottom row}: Vertical component of the magnetic field $B_z$ at $z=0\,\mathrm{km}$, continuum intensity $I$, vertical velocity $v_z$ at $z=700\,\mathrm{km}$, vertical component of the swirling vector $\lambda_z$ at $z=700\,\mathrm{km}$, vertical component of the magnetic swirling vector $\lambda^{B}_z$ at $z=700\,\mathrm{km}$, and the bin-5 intensity $I_5$. Maps of $\lambda_z$ and $\lambda_z^B$ also show the streamlines of the velocity field and the magnetic field projected into the horizontal plane at $z=700\,\mathrm{km}$, respectively.
        }
        \label{fig:SingleSwirlEvent_Appendix}
    \end{sidewaysfigure*}
    Table \ref{tab:ListSupplementarySwirls} lists nine upwardly propagating swirls found in the high-cadence simulation run described in Sect.\,\ref{sec:numerics}.
    This is a non-exhaustive list of events in this time series: We have not attempted to create nor to use an automatized algorithm for detecting swirls and we discard swirls that do not propagate close to the vertical direction. The detection has been carried out manually by looking for chromospheric swirls in the bin-5 intensity $I_5$, over-densities in swirling strength, concentric streamlines in the horizontal velocity field, or vertical propagation of disturbances in magnetic flux concentrations near the $\tau_{500}=1$ surface.  
    All the events listed in Table\,\ref{tab:ListSupplementarySwirls} are unidirectional, upwardly propagating swirling motions, originating from magnetic footpoints near the $\tau_{500}=1$ surface and possessing the characteristics of torsional Alfvén pulses.
    
    Of the nine events of Table\,\ref{tab:ListSupplementarySwirls}, (f) and (c)  were already presented in Figs.\,\ref{fig:SingleSwirlEvent} and \ref{fig:SuperpositionSwirls}, respectively. Another two events, (b) and (e), are depicted in Figs.\,\ref{fig:SuperpositionEvent_Appendix} and \ref{fig:SingleSwirlEvent_Appendix}, respectively. For the rest of events coordinates, time interval, and approximate size are given. The time interval does not necessarily reflect the lifetime of a swirl because it can be affected by a superposition event. Instead, it corresponds to the interval in which the event can be traced in the bin-5 intensity, hence, the interval in which it can be seen in chromospheric layers. The events of Table\,\ref{tab:ListSupplementarySwirls} are marked with a circle in the animation of Fig.\,\ref{fig:Movie_IcoBin5} available online. Figure\,\ref{fig:Movie_IcoBin5} represents the instant $t=5800\,\mathrm{s}$ of that animation. The left panel shows the continuum intensity, the right panel the bin-5 intensity $I_5$. The chosen time instant contains five of the nine events of Table\,\ref{tab:ListSupplementarySwirls}.
\end{appendix}
\end{document}